\documentclass[journal, 12pt,draftcls,onecolumn,twoside]{IEEEtran}
\ifCLASSINFOpdf
\else
\fi
\usepackage{url}
\usepackage[utf8]{inputenc} 
\usepackage[T1]{fontenc}
\usepackage{url}
\usepackage{ifthen}
\usepackage{cite}
\usepackage[cmex10]{amsmath} % Use the [cmex10] option to ensure complicance
\usepackage{cite}
\ifCLASSINFOpdf
 
\else
   \usepackage[dvips]{graphicx}
 \fi
\usepackage[cmex10]{amsmath}
\usepackage{amssymb}
\usepackage{stfloats}
\usepackage{amsthm} 
\usepackage{wrapfig}
\usepackage{amssymb}
\usepackage{latexsym}
\usepackage{bm, bbm} 
\usepackage{epsfig}
\usepackage{graphicx}
\usepackage{epsfig}
\usepackage{multicol}
\usepackage{psfrag}
\usepackage{float}
\usepackage{cite}
\usepackage{subfigure}
\usepackage{color}
\usepackage{balance}

\newcommand{\Z}{\mathbb{Z}}
\newcommand{\Ftwo}{\mathbb{F}_2} 
\newcommand{\code}[1]{\mathcal{#1}}
\newcommand{\matr}[1]{{#1}}
\newcommand{\GF}[1]{\mathbb{F}_{#1}}
\newcommand{\vc}{\vect{c}}
\newcommand{\tr}{\mathsf{T}}
\newcommand{\vect}[1]{\mathbf{#1}}

\newcommand{\defeq}{\triangleq}
\newcommand{\cC}{\mathcal{C}}

\newcommand{\R}{\mathbb{R}}

\newcommand{\codeCQC}[1]{\code{C}_{\mathrm{QC}}^{(N)}}

\newtheorem{lemma}{Lemma}
\newtheorem{theorem}{Theorem}

\newtheorem{corollary}{Corollary}

\theoremstyle{plain}

\newtheorem{PreDefinition}[lemma]{{\textbf{Definition}}}
    {\begin{PreDefinition}}{\hfill$\square$\end{PreDefinition}}

\theoremstyle{plain}

\newtheorem{PreRemark}[lemma]{{\textbf{Remark}}}
  \newenvironment{remark}%
    {\begin{PreRemark}\upshape}{\hfill$\square$\end{PreRemark}}

\newtheorem{PreExample}{{\textbf{Example}}}
  \newenvironment{example}%
    {\begin{PreExample}\upshape}{\hfill$\square$\end{PreExample}}

\usepackage{tikz}

\usepackage{xcolor}

\usepackage{color,soul}

\definecolor{lightblue}{rgb}{.90,.95,1}
\sethlcolor{yellow}

\usepackage{psfrag}

\usepackage[outdir=./]{epstopdf}

\begin{document}

\title{The Number of Cycles  of Bi-regular Tanner Graphs in Terms of the Eigenvalues of the Adjacency Matrix}
\author{Roxana~Smarandache,~\IEEEmembership{Senior Member,~IEEE,} and David~G.~M.~Mitchell,~\IEEEmembership{Senior Member,~IEEE.}% <-this % stops a space
\thanks{This material is based upon work supported by the National Science Foundation under Grant No. CCF-2145917.}% This paper was presented in part at the 2009 IEEE Global Communications Conference \cite{sf09}.}% <-this % stops a space
\thanks{R.~Smarandache is  with the Departments
of Mathematics and Electrical Engineering, University of Notre Dame, Notre Dame, IN 46556, USA (e-mail: rsmarand@nd.edu).}% 
\thanks{D.~G.~M.~Mitchell is with the Klipsch School of Electrical and Computer Engineering, New Mexico State University, NM 88003, USA
(e-mail: dgmm@nmsu.edu).}}% <-this % stops a space
\markboth{IEEE Transactions Information Theory (Submitted paper)}%
{Submitted paper}



\maketitle

% \fi
\IEEEpeerreviewmaketitle
\begin{abstract} In this paper, we explore new connections between the cycles in the graph of low-density parity-check (LDPC) codes and the eigenvalues of the corresponding adjacency matrix. The resulting  observations are used to derive fast, simple, recursive formulas for the number of cycles $N_{2k}$ of length $2k$, $k<g$, in a bi-regular graph of girth $g$. Moreover, we derive explicit formulas for $N_{2k}$, $k\leq 7$, in terms of the nonzero eigenvalues of the adjacency matrix. Throughout, we focus on the practically interesting class of bi-regular quasi-cyclic LDPC (QC-LDPC) codes, for which the eigenvalues can be obtained efficiently by applying techniques used for block-circulant matrices.   % that  can be easily obtained like in my paper with Flanagan
%$$A=\begin{bmatrix} 0&H\\ H^\tr &0\end{bmatrix}.$$ 

   \end{abstract}
\section{Introduction} Counting short cycles in bipartite graphs is a fundamental problem of interest in many fields. For example, the distribution of short cycles that exist in the Tanner graph of a low-density parity-check (LDPC) code largely determines the bit/block error rate performance of the code under iterative decoding algorithms in both the waterfall and error-floor regions  \cite{mb01b,hea05,xb09,aba11,wdw13,bbc18}. 
Motivated by this, significant effort has been made to efficiently compute the cycle distribution of bipartite graphs, see, e.g., \cite{st96,hc05,tee07,kb12,kb13,bl17,db18,db19,db20,db20b,gsm23c}. %These  and to construct graphs that  as well as harmful objects that are composed of cycles, such as trapping sets \cite{kb12b,ehd19,bcb+23}.

Dehghan and  Banihashemi \cite{db20} described a useful connection between the eigenvalues $\{\eta_i\}$ of the directed edge matrix $A_e$ and the eigenvalues $\lambda_i $ of the adjacency matrix $A$ for a connected $(q_1, q_2)$-biregular, bipartite graph $G$ of size $m\times n$.   In particular, for each strictly negative eigenvalue $\lambda$   of $A$, and  for each solution $\xi\neq  1$ of the equation $\xi^2+(-\lambda^2+q_1+q_2) \xi +q_1q_2=0,$  the numbers $\pm\sqrt{\xi}$ are eigenvalues of $A_e$, each taken with the same multiplicity as that of $\lambda$ in the spectrum of $A$. In addition, $\pm\sqrt{-q_1}$  and $\pm\sqrt{-q_2}$, each of  multiplicity 
 $n - \mathrm{rank}(A)/2$ and $m - \mathrm{rank}(A)/2$, respectively, and $\pm 1$, each with multiplicity $|E| - (m + n) + 1$,  are also eigenvalues of $A_e$. These, together with the fact that the number  of $k$-cycles $N_k$ in a bipartite graph $G$ of girth $g$   is given by $N_k = \sum\limits_{i} \frac{\eta_i^k}{2k}$ , for $k < 2g$, gave them a theorem with a  method to  compute $N_{2k}$, $k<g$, starting from the eigenvalues $\lambda$ of $A$.

In this paper, we further explore the steps of this theorem and show how  the computation of $\{\eta_i\}$ from the equation $\xi^2+(-\lambda^2+q_1+q_2) \xi +q_1q_2=0$
can be  bypassed by directly using $\lambda _i$ and by giving a closed formula in which only the powers of the eigenvalues of the matrix $A$ appear, rather than the powers of $\{\eta_i\}$.  We first derive a simple recursive formula for  $N_{2k}$ in a bipartite $(q_1, q_2)$-biregular graph, from which we obtain a formula of $N_{2k}$ dependent  on the sum of $\{\lambda_i^j\}, j\leq k$, and then a formula of $N_{2k}$ as a function of $\sum\{\lambda_i^k\}$ and $N_{2j}, j<k$. 
 
  The paper is structured as follows. In Sections~\ref{definitions} and \ref{background} we give the notations, definitions, and necessary background for the rest of the paper. In  Section~\ref{cycle-theorem}, we list the facts that we use in order to bypass the computation of the powers of the eigenvalues $\eta_i$ of the directed edge matrix and derive the recursive formula dependent on the eigenvalues $\lambda$ of the adjacency matrix $A$.   In Section~\ref {Formula-alpha} we give a formula for $N_{2k}$ in terms of the eigenvalues $\alpha_i =\lambda_i-(q_1+q_2)$, while in Section~\ref {Formula-lambda}  we work out the formula of $N_{2k}$ in terms of the powers of $\lambda$ and, in particular, formulas that depend only on the largest power $\lambda_i^k$ and $N_{2j}, 1<j<k$. We also revisit the examples of Section~\ref{cycle-theorem} to exemplify these formulas. We provide some concluding remarks in Section~\ref{sec:conc} followed by an Appendix in which the proofs of the formulas in  Section~\ref {Formula-lambda} are given. 
  \section{Basic Notation and Definitions}
\label{definitions}
All codes in this paper will be binary linear codes of a certain
length $n$ specified through a (scalar) parity-check matrix
$\matr{H}=(h_{i,j}) \in \GF{2}^{m \times n}$ as the set of all
vectors $\vc \in \Ftwo^n$ \ such that $\matr{H} \cdot \vc^\tr =
\vect{0}^\tr$, where ${}^\tr$ denotes transposition %The minimum
%Hamming distance of a code $\code{C}$ will be denoted by $\dmins(\code{C})$. 
and $\Ftwo$ is the binary field.  For a positive integer $L$, $[L]$ will denote the set of nonnegative integers smaller than $L$, i.e., $[L]=\{0,1,\ldots, L-1\}$. We denote the $N\times N$ identity matrix by $I_N$.

A linear QC-LDPC code $\code{C}_{\rm QC}\defeq\codeCQC{N}$ of length
$n = n_vN$ can be described by an $n_cN \times n_vN$ (scalar) parity-check
matrix $\matr{H}_{\rm QC}^{(N)}\defeq\matr{H}$ that is formed by a $n_c
\times n_v$ array of $N \times N$ circulant matrices
\begin{align}\label{pc-QC}
\matr{H}=\left[
\begin{array}{cccc}
\matr{P}_{1,1}   &       \matr{P}_{1,2}   &       \cdots  &       \matr{P}_{1,n_v} \\
\matr{P}_{2,1}   &       \matr{P}_{2,2}  &       \cdots   &       \matr{P}_{2,n_v}  \\
\vdots &\vdots & &\vdots \\
\matr{P}_{n_c,1}  &       \matr{P}_{n_c, 2}  &    \cdots   &       \matr{P}_{n_c, n_v} 
\end{array}
 \right], 
\end{align}
where the entries $\matr{P}_{i,j}$ are $N\times N$ circulant matrices. Clearly, by choosing these circulant matrices to be low-density, the parity-check matrix will also be low-density. With the help of the well-known isomorphism between the ring of $N\times N$ circulant matrices and the ring of polynomials modulo $x^N - 1$, we can associate a polynomial $p_{i,j}(x)$ to each matrix $\matr{P}_{i,j}$, and thus a QC-LDPC code can equivalently be described by a polynomial parity-check matrix 
\begin{align}\label{eq:matrix_0_bijection}
\matr{H}(x)=\left[
\begin{array}{cccc}
p_{1,1}(x) &   p_{1,2}(x)  &  \cdots  &  p_{1,n_v}(x) \\
p_{2,1}(x)  &   p_{2,2}(x)  &   \cdots   &   p_{2,n_v}(x)  \\
\vdots &\vdots & &\vdots \\
p_{n_c,1}(x)  &   p_{n_c, 2}(x)  &    \cdots &  p_{n_c, n_v}(x) 
\end{array}
 \right],
 \end{align}
of size $n_c\times n_v$, with polynomial operations performed modulo $x^N-1$. 

By permuting the rows and columns of the scalar parity-check matrix
$\matr{H}$,\footnote{This permutation is performed by taking the first row in the first block of $N$ rows, the first row in the second block of $N$ rows, etc., then the second row in the first block, the second row in the second block, etc., and then similarly for the columns.} we obtain an equivalent
parity-check matrix representation for the QC code $\codeCQC{N}$ as
\begin{align} \matr{\bar H}
    &\defeq
       \begin{bmatrix}
         \matr{H}_0      & \matr{H}_{N-1}      & \cdots 
                         & \matr{H}_1 \\ 
         \matr{H}_1      & \matr{H}_0          & \cdots   
                         & \matr{H}_2 \\ 
         \vdots          & \vdots              & \ddots           
                         & \vdots \\ 
                  \matr{H}_{N-1}   & \matr{H}_{N-2} & \cdots 
                         & \matr{H}_0
  \end{bmatrix}, 
\label{eq:matrix_1_bijection}
\end{align}
where $\matr{H}_0, \matr{H}_1, \ldots, \matr{H}_{N-1}$ are scalar $n_c
\times n_v$ matrices and the connection between the two representations is
\begin{align}
\matr{H}_0 + \matr{H}_1 x + \cdots + \matr{H}_{N-1} x^{N-1}=\matr{H}(x).
\label{eq:matrix_2_bijection}
\end{align}

% A linear QC-LDPC code $\code{C}_{\rm QC}\defeq\codeCQC{N}$ of length
%$n = n_vN$ can be described 
%%by an $rJ \times rL$ (scalar) parity-check
%%matrix $\matr{\bar H}_{\rm QC}^{(r)}\defeq\matr{\bar H}$ that is formed by a $J
%%\times L$ array of $r \times r$ circulant matrices $\matr{\bar H}$, or, equivalently,
% by
%a polynomial parity-check matrix $\matr{P}(x)$ of size $n_c \times  n_v$,
%with polynomial operations performed modulo $x^N-1$:
%\begin{align}\label{matrix-P}
%%\matr{\bar H}=\left[
%%\begin{array}{cccc}
%%\matr{P}_{1,1}   &       \matr{P}_{1,2}   &       \ldots  &       \matr{P}_{1,L} \\
%%\matr{P}_{2,1}   &       \matr{P}_{2,2}  &       \ldots   &       \matr{P}_{2,L}  \\
%%\vdots &\vdots &\ldots &\vdots \\
%%\matr{P}_{J,1}  &       \matr{P}_{J, 2}  &    \ldots   &       \matr{P}_{J, L} 
%%\end{array}
%% \right], 
% \matr{P}(x)=\left[
%\begin{array}{cccc}
%p_{1,1}(x)   &       p_{1,2}(x)  &       \ldots  &       p_{1,n_v}(x) \\
%p_{2,1}(x)   &       p_{2,2}(x)  &       \ldots   &       p_{2,n_v}(x)  \\
%\vdots &\vdots &\ldots &\vdots \\
%p_{n_c,1}(x)  &       p_{n_c, 2}(x)  &    \ldots   &       p_{n_c, n_v}(x) 
%\end{array}
% \right].
%\end{align}
%where the entries $\matr{P}_{i,j}$ are $N\times N$ circulant matrices.

For any integer $N \ge 1$, let $R_N\defeq  \{ e^{i 2 \pi j / N} ,  0 \le j < N \}$
denote the set of complex $N$-th roots of unity.  %and let $R_s^{-} =
%R_s\backslash \{ 1 \}$. 
The symbol ${}^*$ denotes complex conjugation. An $Nn_c\times Nn_c$ circulant matrix $\matr{B}$, whose entries are an $N\times N$ array of square $n_c \times n_c$ matrices, will be called an \emph{$n_c$-block circulant matrix}; we shall denote this by
$\matr{B} = \mathrm{circ}(\vect{b}_0, \vect{b}_1, \ldots, \vect{b}_{N-1})\in \cC^{Nn_c\times Nn_c}$
where the square $n_c \times n_c$ matrix entries in the first block column of $\matr{B}$ are $\vect{b}_0, \vect{b}_1,\ldots , \vect{b}_{N-1}$, respectively. 
% If $\matr{B}\in \cC^{N\times N}$ is a
%circulant matrix and $w(x)= b_0+b_1x+\ldots +b_{N-1}x^{N-1}$ its
%(column) associated polynomial, then the eigenvalues of $\matr{B}$ are
%given by this polynomial's evaluation $w(\rho^j)$ at the complex $N$-th roots of unity $\rho^j$, $\rho =e^{i\frac{2\pi j}{N}} \in R_N, i^2=-1$, for $j=0,1,\ldots , N-1$,  \cite{}[{MacWilliams:Sloane:98}], 
%with associated eigenvector $\vect{v}\defeq \begin{bmatrix} 1& \rho^j& \rho^{2j}&\ldots & \rho^{(N-1)j}\end{bmatrix}^\tr$. 
%The following gives a proof of this result based on the polynomial representation of a circulant
%matrix. It may be seen as a special case of the method we present later for QC codes.
%
%Let  $\lambda$ be an eigenvalue of $\matr{B}$. Then there exists a nonzero
%vector $\vect{v}=(v_0, \ldots,
% v_{n-1})^\tr \in \cC^{n}$ such that
%\begin{align*}
%&\matr{B}\vect{v}=\lambda\vect{v}.
% \end{align*}
% In polynomial form, this equation is equivalent to (here $v(x) = v_0+v_1x+\ldots +v_{n-1}x^{n-1}$):
% \begin{align*}
%   &w(x)v(x)=\lambda v(x) \mod (x^n-1) {~\rm iff}
%   \\& x^n-1~|~w(x)v(x)-\lambda v(x) {~\rm in~} \cC{~\rm iff}\\
%   &w(x)v(x)=\lambda v(x), \forall x \in R_n{~\rm iff}\\
%   &(w(x)-\lambda)v(x)=0, \forall x\in R_n \; .\\
%%&\lambda=c(\rho^i)v(\rho^i)/v(\rho^i)=c(\rho^i)
%\end{align*}
%For each $x\in R_n$, $\lambda= w(x)$ is a solution of the above
%equation, and therefore it is an eigenvalue for the matrix $\matr{B}$.
%There are $n$ such solutions, therefore, these are all possible
%eigenvalues of $\matr{B}$.
If  %$\matr{B}=\mathrm{circ}(\vect{b}_0, \vect{b}_1, \cdots, \vect{b}_{N-1})\in \cC^{Nn_c\times Nn_c}$ is  an \emph{$n_c$-block circulant} matrix,  and 
 $\matr{W}(x)= \vect{b}_0+\vect{b}_1x+\cdots
  +\vect{b}_{N-1}x^{N-1}$ is the associated (column) matrix polynomial of $B$, then the eigenvalues of $\matr{B}$ are given by the union of the
  eigenvalues of the $n_c\times n_c$ matrices $\matr{W}(\rho)$, for all $\rho\in
  R_N$~\cite{tee07}. In particular, if $\vect{v}$ is an eigenvector for a given  eigenvalue $\lambda$ of  $\matr{W}(\rho)$, for some $\rho$ in $R_N$, then the vector 
  $\begin{bmatrix} \vect{v}^\tr& \rho\vect{v}^\tr& \rho^{2}\vect{v}^\tr&\cdots & \rho^{(N-1)}\vect{v}^\tr\end{bmatrix}^\tr$ is an eigenvector for $\matr{B}$. 
 
  \section{Linear Algebra background} \label{background} 
   We briefly remind the reader of some necessary results to make this paper self-contained.
\subsection{Eigenvalues}\label{eigenvalues} 
\begin{itemize} 
\item Let $M$ be an   $m\times n$ matrix. Then the $m\times m$ matrix  $MM^T$ and the $n\times n$ matrix $M^T M$ have the same eigenvalues (all real), except a set of $|m-n|$ zero-eigenvalues that make up for the difference $|m-n|$ in matrix sizes. 
 
\item ${\rm rank}(MM^T)= {\rm rank}(M^TM)={\rm rank}(M)={\rm rank}(M^T) =\# \{\text{nonzero eigenvalues of} MM^T\}$.
\item Let $p(t)=a_n t^n+a_{n-1}t^{n-1}+\cdots+ a_1t+a_0$ be any polynomial of degree $n$,  $M$ an $m\times m$ matrix, and  $\lambda$ any eigenvalue of  $M$  with an eigenvector $\vect{x}$. Then $p(\lambda)\defeq a_n\lambda^n +a_{n-1}\lambda^{n-1}+\cdots+ a_1\lambda+a_0 $ is an eigenvalue for $p(M)\defeq a_n M^n+a_{n-1}M^{n-1}+\cdots+ a_1M+a_0I_{n}$ with the same eigenvector $\vect{x}$. 

\begin{itemize} \item In particular,  the eigenvalues of $M^k$ are $\lambda^k$, for all $k=1, 2, \ldots$. \end{itemize} \end{itemize}
\subsection{Newton's Identities} 
\begin{itemize}\item If $p(t)= a_n t^n+a_{n-1}t^{n-1}+\cdots+ a_1t+a_0$  is a  monic polynomial, i.e., $a_n=1$, and has  zeros $\alpha_1, \ldots, \alpha_n$,   including  multiplicities, and  $E_k\defeq \sum\limits_{i=1}^n \alpha_i^k, k=1,2, \ldots , $  are the $k$th {\it moments} of the zeros, then Newton's identities hold: \begin{itemize}  
\item for all $k=1,2,\ldots, n,$
  \quad $E_{k} +a_{n-1}E_{k-1}+a_{n-2}E_{k-3} +\cdots +  a_{n-k+1}E_{1} + a_{n-k}k=0,$
  \item   for $k=1,2,\ldots, $ \quad 
   $E_{n+k} +a_{n-1}E_{n+k-1}+a_{n-2}E_{n+k-3} +\cdots +  a_0 E_{k}=0.$
 \end{itemize}
\item  In particular, the $k$th moments  $E_k\defeq \sum\limits_{i=1}^m \lambda_i^k$, $k=1,2, \ldots $, of the eigenvalues $\lambda_1, \ldots, \lambda_m$,  including  multiplicities, of an $m\times m$ real matrix $M$, can be computed recursively as above using the (monic) characteristic polynomial  $p(t)=\det(t I_m-M)$ of $M$,   which has roots at the eigenvalues $\lambda_1, \ldots, \lambda_m$.
% for $k=1,2,\cdots, n$, 
%  $$E_{k} -b_{n-1}E_{k-1}+b_{n-2}E_{k-3} -\ldots + (-1)^{n-k+1} b_{n-k+1}E_{1} +(-1)^{n-k} b_{n-k}k=0$$
%  and for $k=1,2,\cdots$, 
%   $$E_{n+k} -b_{n-1}E_{n+k-1}+b_{n-2}E_{n+k-3} -\ldots + (-1)^n b_0 E_{k}=0.$$ 
%\end{lemma}

 \end{itemize}

 \subsection{The Adjacency Matrix $A$ of a Tanner Graph Associated with a Parity-check Matrix $H$} 
 Suppose that $H$ is  an $m\times n$ matrix, $m<n$.%\footnote{If $m>n$, we substitute $ HH^\tr$ above by $H^\tr H$.} 
  
 \begin{itemize} \item The adjacency matrix $A$ of a Tanner graph associated with a parity-check matrix $H$ is $$A\defeq \begin{bmatrix} 0&H\\ H^\tr &0\end{bmatrix}\Longrightarrow  A^2=\begin{bmatrix} HH^\tr&0\\0&  H^\tr H \end{bmatrix}.$$  
  \item  The following are equivalent:
  \begin{enumerate} \item $\lambda \neq 0$ is an eigenvalue of $HH^\tr$ of multiplicity $\mu$;\item  $\lambda \neq 0$ is an eigenvalue of $A^2$ of multiplicity $2\mu$ ; 
  \item  Nonzero $-\sqrt{\lambda}, \sqrt{\lambda}$,  are  eigenvalues of $A$, each of multiplicity $\mu$;
  \item  $ \sqrt{\lambda}\neq 0$  are  singular values  of $H$ of multiplicity $\mu$.  
   \end{enumerate} 
  \item The rank of $A$  is twice the rank of $H$.    
  \end{itemize}
%\begin{lemma}\label{Newton}
 
  \subsection{The Eigenvalues of $\matr{H}^{\tr}\matr{H}$ 
for a QC Code}
\label{sec:eigenvalues}
 We recall, prove, and exemplify a previous result concerning the eigenvalues of QC-LDPC codes.

\begin{corollary} [\!\!\cite{sf09}] \label{cor1} Let $\code{C}_{\rm QC}\defeq\codeCQC{N}$ be a  QC-LDPC code with  parity-check
matrices $\matr{H}, \matr{H}(x), {\bar H}$ as in \eqref{pc-QC}, \eqref{eq:matrix_0_bijection} and \eqref{eq:matrix_1_bijection}. Then the eigenvalues of $\matr{H}\matr{H}^\tr$ are given by the union of the eigenvalues of the $n_c\times n_c$ matrices $\matr{H}(\rho)\matr{H}(\rho)^\tr,$ for all $N$th roots of unity $\rho =e^{i\frac{2\pi j}{n}} $ in $R_N, i^2=-1$, for $j=0,1,\ldots , N-1$.
\end{corollary} 
\begin{IEEEproof} For a fixed value of $N \ge 1$, \eqref{eq:matrix_1_bijection} and  \eqref{eq:matrix_2_bijection} provide a simple bijective correspondence between the set of polynomial matrices $\matr{H}(x) \in (\R[x]/(x^N-1))^{n_c \times n_v}$ and the set of parity-check matrices $\matr{\bar H}$. Furthermore, the product of two such polynomial matrices $\matr{H}(x)\matr{L}(x)$, where defined, yields another correspondence  via this bijection with the product of the corresponding parity-check matrices $\matr{\bar H}\matr{\bar L}$, while the transposition $\matr{H}(x)^\tr$ in the form of \eqref{eq:matrix_0_bijection} corresponds to transposition of the corresponding parity-check matrix in the form of $\matr{\bar H}^\tr$. In particular,  $\matr{H}(x)\matr{H}(x)^\tr \in (\R[x]/(x^N-1))^{n_c \times n_c}$ 
corresponds to the $n_c$-block circulant matrix ${\bar H}{\bar H}^\tr$, with the eigenvalues of $\matr{H}(\rho)\matr{H}^\tr(\rho)$, for all $\rho\in R_N$, see  Section \ref{definitions}. 
\end{IEEEproof}

\begin{example}\label{Tannerexample}
  Let $N=31$ and consider the $(3,5)$-regular $[155, 91, 20]$ QC-LDPC code~\cite{tss+04} given by the
   $93 \times 155$ scalar matrix \begin{align*}
  {\matr{ H}} &= \begin{bmatrix}
    \matr{P}_1   & \matr{P}_2    & \matr{P}_4    & \matr{P}_8 & \matr{P}_{16}\\
    \matr{P}_5   & \matr{P}_{10} & \matr{P}_{20} & \matr{P}_9 & \matr{P}_{18}\\
    \matr{P}_{25}& \matr{P}_{19} & \matr{P}_7 & \matr{P}_{14}&
    \matr{P}_{28}
         \end{bmatrix},
         \end{align*}
where $\matr{P}_\ell$ denotes the $31 \times 31$ identity matrix with rows shifted cyclically to the left by $\ell$ positions or, equivalently, by the polynomial parity-check matrix 
\begin{align*}\matr{H}(x)
       = \begin{bmatrix}
           x      & x^2    & x^4    & x^8 & x^{16}\\
           x^5    & x^{10} & x^{20} & x^9 & x^{18}\\
           x^{25} & x^{19} & x^7    & x^{14}& x^{28}
         \end{bmatrix}, \;  \matr{H}^\tr(x) = \begin{bmatrix}
    x^{30}      & x^{26} &x^{6} \\  x^{29} & x^{21} & x^{12} \\     x^{27}   & x^{11} & x^{24}\\  x^{23}  & x^{22}& x^{17}\\  x^{15}& x^{13} &  x^{3}
         \end{bmatrix},   
\end{align*} 
where $\matr{H}(x)\in (\R[x]/(x^{31}-1))^{3 \times 5}$.

The eigenvalues of the associated  scalar matrix $HH^T$  are computed by evaluating 
  %\end{align*}
  %
%  The corresponding matrix
%  $\matr{H}$ in the form (\ref{eq:matrix_1_bijection}) is a $31\times 31 $ matrix with block entries
%  $\matr{H}_i$, $i\in [31]$ obtained by decomposing $\matr{P}(X)$
%  according to the powers of $X$:
%  \begin{align}
%\matr{P}(X)=\matr{H}_0 + \matr{H}_1 X + \cdots + \matr{H}_{30} X^{30}.
%\end{align}
%Obviously only $15$ matrices among the $\matr{H}_i$ are nonzero, and all
%of these contain only one $1$, the other entries being zero.
%
%The matrix $\matr{H}^\tr\cdot\matr{H}$ is a $5$-block circulant matrix. Corollary \ref{cor:QC_codes} above tells us
%that in order to compute its eigenvalues, we need to form the matrices
%$\matr{P}^\tr(\rho^{-i})\cdot \matr{P}(\rho^i)$, for all $i\in [31]$ (here $\rho$ denotes a primitive complex $31$-th root of unity). We
%have that
%\begin{align*}
%  \matr{P}^\tr(1/x) &= \begin{bmatrix}
%    x^{30}      & x^{29}    & x^{27}    & x^{23} & x^{15}\\
%    x^{26}    & x^{21} & x^{11} & x^{22} & x^{13}\\
%    x^{6} & x^{12} & x^{24} & x^{17}& x^{3}
%         \end{bmatrix}^\tr \; \quad 
     \begin{align*}  \matr{H}(x)\matr{H}^\tr(x)=\begin{bmatrix}
    5  & f_1^\tr & f_2 \\
    f_1 & 5 &f_3^\tr  \\
    f_2^\tr& f_3 & 5  \end{bmatrix} \end{align*} with
    \begin{align*}
    f_1 & \defeq x+ x^{2}+x^{4} +  x^{8} + x^{16}, \\
    f_2 & \defeq x^{25} +  x^{19} + x^{7} + x^{14}+ x^{28}, \\
     f_3 & \defeq  x^{5} +  x^{10} + x^{20} + x^{9}+ x^{18},   
%    
%     \matr{P}^\tr(1/x) \matr{P}(x)=
%  \begin{bmatrix}
%    3  & a & e^* & c  & e^* \\
%    a^* & 3 & b & a^* & d \\
%    e & b^* & 3 & c & b^* \\
%    c^* & a & c^* & 3 & d \\
%    e & d^* & b & d^* & 3 \end{bmatrix} \;, 
 \end{align*}
%    
%     \matr{P}^\tr(1/x) \matr{P}(x)=
%  \begin{bmatrix}
%    3  & a & e^* & c  & e^* \\
%    a^* & 3 & b & a^* & d \\
%    e & b^* & 3 & c & b^* \\
%    c^* & a & c^* & 3 & d \\
%    e & d^* & b & d^* & 3 \end{bmatrix} \;, 
% \end{align*}
at  $\rho^k =\exp\left(i\frac{2 \pi k }{ 31}\right)$, for all $k=0,1,\ldots, 30$, where $f_i^\tr(x) \defeq f_i(x^{-1})$ and all exponents are taken modulo $N=31$.  
%\begin{align*}
%  a&=\rho + \rho^5 + \rho^{25}; b = \rho^2 + \rho^{10} + \rho^{19};
%  c = \rho^4 + \rho^{20} + \rho^7;\\
%  d& =\rho^8 + \rho^9 + \rho^{14}; e = \rho^{16} + \rho^{18} +
%  \rho^{28}.
%\end{align*}
%\begin{align*} f_1\defeq & x^{4} +  x^{8} + x^{16} + x^{1}+ x^{}, f_2\defeq & x^{4} +  x^{8} + x^{16} + x^{1}+ x^{};  &c = x^4 + x^7 + x^{20};\\
%  d& =x^8 + x^9 + x^{14}; e = x^{16} + x^{18} +
%  x^{28}.
%  a&=x + x^5 + x^{25}; b = x^2 + x^{10} + x^{19};
%  c = x^4 + x^7 + x^{20};\\
%  d& =x^8 + x^9 + x^{14}; e = x^{16} + x^{18} +
%  x^{28}.
%\end{align*}
For all $k\in [31], k\neq 0$,  $
\matr{H}(\rho^k)\matr{H}^\tr(\rho^{k})$ has  characteristic  polynomial %of $\matr{P}^\tr(\rho^{i})\cdot
%\matr{P}(\rho^i)$ 
\begin{align*}
p(t) &\defeq \det\left(tI-\matr{H}(\rho^k)\matr{H}^\tr(\rho^{k})\right) = t^3 - 15 t^2 + 62 t - 62 \\&=
 (t-8.6801)(t-4.8459)(t-1.4740),
\end{align*}
%  Note that the coefficients of the characteristic polynomials can be computed from the principal minors of $H(x)H(x)^\tr$ as following:
% \begin{itemize} \item  its trace is equal to the negative of the coefficient of $t^2$ evaluated at all $\rho^j $:
% $$\sum_{j=1}^3 (H(x)H(x)^\tr)_{j,j} =5+5+5=15$$ 
%\item The coefficient of $t$ is the sum of all 2 by 2 principal minors evaluated at all $\rho^j $
%$$62=\left| \begin{matrix} 5  & f_1^\tr(\rho^j) \\
%    f_1(\rho^j) & 5  \end{matrix} \right| + \left| \begin{matrix} 5  & f_2(\rho^j) \\
%    f_2^\tr (\rho^j)& 5  \end{matrix} \right| + \left| \begin{matrix} 5  & f_3^\tr(\rho^j)\\
%    f_3(\rho^j) & 5  \end{matrix} \right| =75-\sum_{j=1}^3 f_jf_j^\tr(\rho^j) =62$$
%\item the  coefficient of $t^0$ is the negative of the sum of all $3\times 3$ principal minors, which is just the determinant of $H(x)H(x)^\tr$ evaluated at all $\rho^j$.
% {\small \begin{align*}&\left| \begin{matrix} 
% 5  & f_1^\tr(\rho^j) & f_2(\rho^j) \\
%    f_1(\rho^j) & 5 &f_3^\tr(\rho^j) \\
%    f_2^\tr(\rho^j)& f_3(\rho^j)& 5  \end{matrix} \right| =125+(f_1f_2f_3+f_1^\tr f_2^\tr f_3^\tr)(\rho^j) -5\sum_{j=1}^3 f_jf_j^\tr(\rho^j)=\\&125+(f_1f_2f_3+f_1^\tr f_2^\tr f_3^\tr)(\rho^j) -5\cdot13=60 +(f_1f_2f_3+f_1^\tr f_2^\tr f_3^\tr)(\rho^j) \end{align*}}   
%    \end{itemize} 
leading to eigenvalues $ 8.6801, 4.8459$ and $1.4740$, each with multiplicity  $30$. This high multiplicity arises a result of the structure of the matrix $H(x)$. 

Indeed, the rows $H_k(x)$ of $H(x)$ satisfy $H_1(x^{25})= H_2(x^5)=H_3(x)$,  and since $2^5=1$ and $5^3=1$ in $\Z_{31},$ we have that 
$$f_j(\rho^k)=f_j(\rho^{2k})=f_j(\rho^{4k})=f_j(\rho^{8k})= f_j(\rho^{16k}), \text{ for } j=1,2,3, \text{ and }k\in [31], $$
and, for all $k\in [31]$, 
$$f_1(\rho^{25k})=f_2(\rho^{k})=f_3(\rho^{5k}). $$  
Therefore, when the roots of unity $\rho^j$, $1\leq j\leq 30$,  are evaluated in place of  $x$ in  $H(x)H(x)^\tr$, the matrix  obtained is either equal to $\matr{H}(\rho)\matr{H}^\tr(\rho)$, for $j=1,2,4,8,16$, or to a matrix obtained from it by symmetric row and column permutations (i.e., by performing  the same column operations as the row operation performed), where $$\matr{H}(\rho)\matr{H}^\tr(\rho)=\begin{bmatrix} 
5&  1.5419 - 2.2180i & -0.3934 - 0.8726i \\
   1.5419 + 2.2180i  & 5& -1.6485 - 1.4384i\\
  -0.3934 + 0.8726i & -1.6485 + 1.4384i &  5\end{bmatrix}.$$ 
  Therefore, $H(x)H(x)^\tr$ evaluated at $\rho^j$, for $1\leq j\leq 30$, has the same 3 eigenvalues, $ 8.6801$, $4.8459$, and $1.4740$.
 At  $\rho^0=1$,  the matrix 
$\matr{H}(1)\cdot \matr{H}^\tr(1)=5J_3$, where $J_3$ is the $3\times 3$ all-one matrix, so it has eigenvalues $15,0,0$.  
This yields the nonzero eigenvalues of $HH^\tr $ as $15$,  $8.6801$, $4.8459$, and $1.4740$ with multiplicities $1$, $30$, $30$, and $30$, respectively, and hence the rank of $H$ and $HH^\tr$ are equal to $91$.
\end{example} 

%\begin{remark} not true A $(q_1, q_2)$- biregular $n_c N\times n_vN$ protograph-based matrix, there are at most $n_c$  nonzero eigenvalues, {\it possibly equal},  each with multiplicity $N-1$, together with the eigenvalue $(q_1+1)(q_2+1) $ of multiplicity 1. In Example~\ref{Tannerexample}, we have $n_c=3$ distinct nonzero eigenvalues each of multiplicity $N-1=30$, together with  the eigenvalue $(q_1+1)(q_2+1)=15$ of multiplicity 1, while in the later Example~\ref{example2}, we will see that the nonzero eigenvalues are $n, n, n$, each of multiplicity $(N-1)=(n-1)$, while $(q_1+1)(q_2+1)=3n$ has multiplicity 1, therefore, it has distinct eigenvalues $n$ and $3n$ of multiplicities $3(n-1)$ and 1, respectively.  
%\end{remark} 
\section{Counting Cycles}\label{cycle-theorem} 
\subsection{Existing Results}
Before further developing our theory, we  begin this section by recalling two important theorems. The first shows how to  compute the number of cycles in a bipartite biregular graph using the spectrum of the directed edge matrix~\cite{kb12}, while the second shows  how this spectrum is connected to the spectrum of the adjacency matrix~\cite{db20}.

\begin{theorem} [\!\!\cite{kb12}] 
\label{thm1}
Consider a bipartite graph $G$ with the directed edge matrix  $A_e$,\footnote{The directed edge matrix is a $2|E|\times 2|E|$ matrix that characterizes the $|E|$ edges in the bipartitie graph. For a definition of the directed edge matrix, we refer the reader to~\cite{kb12}.}  and let $\{\eta_i\}$ 
be the spectrum of $A_e$. Then, the number of $k$-cycles in $G$ is given by $N_k = \sum\limits_{i} \frac{\eta_i^k}{2k}$ , for $k < 2g$, where $g$ is the girth of $G$. \end{theorem}

\begin{theorem}[\!\!\cite{db20}] 
\label{thm2}
 Let $G = (V = U \cup  W, E)$  be a connected bi-regular bipartite graph such that each node in $U$ has degree $q_1 +1$  and each node in $W$ has degree $q_2 +1$, where $q_2 \geq 2, q_1 \geq 1$  and $q_2 \geq  q_1$. Also, assume that $|U| = n$ and $|W| = m$. The eigenvalues of the directed edge matrix $A_e$ of $G$ can then be computed from the eigenvalues of the adjacency matrix A as follows: 
 
 Step 1. For each strictly negative eigenvalue $\lambda$   of $A$, use equation $$\xi^2+(-\lambda^2+q_1+q_2) \xi +q_1q_2=0$$ to find two solutions. For each solution $\xi\neq  1$, the numbers $\pm\sqrt{\xi}$ are eigenvalues of $A_e$, each with the same multiplicity as that of $\lambda$ in the spectrum of $A$. (The total number of eigenvalues of $A_e$ obtained in this step is $2(m+n)-2{\rm nullity}(A)-2$.)

Step 2. Matrix $A_e$ also has the eigenvalues $\pm\sqrt{-q_1}$  and $\pm\sqrt{-q_2}$.\footnote{ Note that $\pm\sqrt{-q_1}, \pm\sqrt{-q_2}$ are solutions of $\xi^2+(-\lambda^2+q_1+q_2) \xi +q_1q_2=0$ for $\lambda= 0$.} The multiplicity of each of
the eigenvalues $\pm\sqrt{-q_1} $ (resp. $\pm\sqrt{-q_2}$)  is $n - {\rm rank}(A)/2$ (resp. $m - {\rm rank}(A)/2$). (The total number of of eigenvalues of $A_e$  obtained in this step is $2(m + n) -2{\rm rank}(A) = 2{\rm nullity}(A)$.)

Step 3. Furthermore, matrix $A_e$ has eigenvalues $\pm 1$, each with multiplicity $|E| - (m + n) + 1$. 
(The total number of eigenvalues in this step is $2|E| - 2(m + n) + 2$.)
\end{theorem}
%\subsection{Formula derivation for the number of cycles}
\subsection{Observations}\label{sec:obs}
We will later rewrite the formula from Theorem \ref{thm1} using the following observations:
 \begin{enumerate} 
  \item For a bipartite graph, only cycles of even lengths exist, so the sums of interest are  
 $$N_{2k}=\sum\limits_{\eta} \frac{\eta^{2k}}{4k}= \sum\limits_{\eta} \frac{(\eta^{2})^{k}}{4k};$$
 \item  The following are equivalent: \begin{enumerate} \item Nonzero  $\pm \lambda$  are eigenvalues of $A$, each  of  multiplicity $\mu_\lambda$;  
 \item $\lambda^2$ is an eigenvalue of $HH^\tr$ of multiplicity $\mu_\lambda$; 
 \item $\alpha\defeq \lambda^2-(q_1+q_2)$ is an eigenvalue of $HH^\tr -(q_1+q_2)I_{m}$ of multiplicity $\mu_\alpha=\mu_\lambda$; \end{enumerate}
 
   \item  Let  $\pm\sqrt{\xi_{\alpha,1}}, \pm\sqrt{\xi_{\alpha,2}}$ be   eigenvalues of  $A_e$, where $\xi_{\alpha,1}, \xi_{\alpha,2}$ are roots of  $$\xi^2-\alpha \xi +q_1q_2,$$ with $\alpha=\lambda^2-(q_1+q_2)$  an eigenvalue of $HH^\tr -(q_1+q_2)I_{m}$ of multiplicity $\mu_\alpha$, and such that $\alpha\neq -(q_1+q_2)$.  %.$\lambda_i^2$ an eigenvalue of $HH^\tr$ of multiplicity $\mu_i$, . 
   Then \begin{align*}&\frac{\sum\limits _{\alpha}\mu_\alpha(+\sqrt{\xi_{\alpha,1}})^{2k}+\mu_\alpha(-\sqrt{\xi_{\alpha,1}})^{2k}+ \mu_\alpha(+\sqrt{\xi_{\alpha,2}})^{2k}+\mu_\alpha(-\sqrt{\xi_{\alpha,2}})^{2k}}{4k}\\&=\frac{\sum\limits _{\alpha}\mu_\alpha(\xi_{\alpha,1}^k +\xi_{\alpha,2}^k)}{2k},\end{align*}
    where the sum is over all {\em distinct } eigenvalues of $HH^\tr-(q_1+q_2)I_{m}$ 
   %So this  sum is indexed   over all {\it distinct} nonzero eigenvalues of $HH^\tr$, or, equivalently, over all eigenvalues $\alpha_i, i=1,2\ldots, l$, of $HH^\tr -(q_1+q_2)I_{m}$ 
   not equal to $-(q_1+q_2)$, hence the multiplicity  $\mu_\alpha$ of each appears in the sum;  
   
\item   Equivalently,  indexing the sum over  all eigenvalues $\alpha\neq -(q_1+q_2)$ of $HH^\tr -(q_1+q_2)I_{m}$, there are $\mu =\sum\limits_{\alpha\neq -(q_1+q_2)}\mu_\alpha={\rm rank}(H)$  such  eigenvalues (not necessarily distinct). The sum is equal to 
         $$\frac{\sum\limits _{{{\small\text{distinct }}\alpha}\atop {\alpha\neq -(q_1+q_2)} }\mu_\alpha(\xi_{\alpha,1}^k +\xi_{\alpha,2}^k)}{2k}=\frac{\sum\limits _{{\alpha}\atop {\alpha\neq -(q_1+q_2)}} (\xi_{\alpha,1}^k +\xi_{\alpha,2}^k)}{2k};$$
   \item Since $\pm\sqrt{-q_1}$  and $\pm\sqrt{-q_2}$  are the  eigenvalues of  $A_e$, then $(\pm\sqrt{-q_1})^{2k}=(-q_1)^k$  and $(\pm\sqrt{-q_2})^{2k}=(-q_2)^{k}$, and they have the multiplicity $n-\mu$ and $m-\mu$, where rank$(H)=\mu$. Therefore,   
\begin{align*}&\frac{((\sqrt{-q_1})^{2k} +(-\sqrt{-q_1})^{2k}) ( n-\mu) +((\sqrt{-q_2})^{2k} +(-\sqrt{-q_2})^{2k} ) (m-\mu)}{4k} \\&=\frac{(-q_1)^k ( n-\mu ) +(-q_2)^{k} (m-\mu)}{2k}; \end{align*}
   \item Since $\pm 1$ are the  eigenvalues of  $A_e$ of multiplicity $|E| -(m+n)+1$, then 
   $$\frac{(-1)^{2k}(|E| -(m+n))+1^{2k} (|E| -(m+n)+1}{4k}=\frac{ |E| -(m+n)+1}{2k}.$$
 Also,  $|E|=n(q_1+1)= m(q_2+1)$. 
 \item For the eigenvalue $\alpha_0=(q_1+1)(q_2+1)- (q_1+q_2)= q_1q_2+1$, the roots of $\xi^2-\alpha_0 \xi +q_1q_2=0$ are $1$ and $q_1q_2$, of multiplicity 1. Therefore, 
$$\frac{%\sum\limits _{\alpha}\mu_0
(\xi_{0,1}^k +\xi_{0,2}^k)}{2k} =\frac{1+q_1^kq_2^k }{2k} .$$ However, the eigenvalues equal to  $1$ are already computed in $|E| -(m+n)+1$, so we need to subtract $1$ from $|E| -(m+n)+1$, so 
  we have the two sums together:
  { \small $$\frac{ |E| -(m+n)+\sum\limits _{{\alpha}\atop {\alpha\neq -(q_1+q_2)}} (\xi_{\alpha,1}^k +\xi_{\alpha,2}^k)}{2k}=
   \frac{ |E| -(m+n)+1+(q_1q_2)^k+\sum\limits _{{\alpha}\atop {\alpha\neq -(q_1+q_2), q_1q_2+1}} (\xi_{\alpha,1}^k +\xi_{\alpha,2}^k)}{2k};$$}

% therefore,  we either add  $\frac{q_1^kq_2^k }{2k} $ into the $N_{2k}$ sum, or subtract $1$ from $|E| -(m+n)+1$, and consider $\alpha_0$ part of the sum $S_{\alpha, k}\defeq \xi_{\alpha, 1}^k +\xi_{\alpha,2}^k$ computed below. 
%  
  \item For each $\alpha $  eigenvalue of $HH^\tr-(q_1+q_2)I_m$, $\alpha \neq -(q_1+q_2)$, we compute the sums $S_{\alpha, k}\defeq \xi_{\alpha, 1}^k +\xi_{\alpha,2}^k$ recursively as functions of powers of $\alpha$, using Newton's identities, as follows. 
   Since $\xi_{\alpha,1}, \xi_{\alpha,2}$  are roots of $\xi^2-\alpha\xi +q_1q_2,$   then $$\xi_{\alpha,1}+ \xi_{\alpha,2}= \alpha, \quad \xi_{\alpha,1} \xi_{\alpha,2}= q_1q_2,$$ therefore, 
\begin{align}\label{S-functions}
  & S_{\alpha, 1}\defeq \alpha, S_{\alpha, 2} =S_{\alpha, 1}^2-2q_1q_2=\alpha^2-2q_1q_2, % \nonumber\\ 
    S_{\alpha, k}= %S_{\alpha, k-1} S_{\alpha, 1}-q_1q_2 S_{\alpha, k-2} = 
    \alpha S_{\alpha, k-1}-q_1q_2 S_{\alpha, k-2}, k\geq 3. 
   \end{align}
We obtain
 \begin{align*}
 &S_{\alpha, 1}= \alpha , %\\&
  S_{\alpha, 2}=%\defeq S_{1,i}^2- 2q_1q_2=
   \alpha^2- 2q_1q_2, %\\&
S_{\alpha, 3}= \alpha^3- 3q_1q_2\alpha, %\\&
  % \alpha_i(\alpha_i^2-3q_1q_2) \\  
  S_{\alpha, 4} =%\defeq S_{2,i}^2- 2q_1^2q_2^2=(\alpha_i^2- 2q_1q_2)^2- 2q_1^2q_2^2=
   \alpha^4- 4q_1q_2\alpha^2 + 2q_1^2q_2^2\\ 
   &S_{\alpha, 5}= %\defeq S_{1,i}^5- {5\choose 1} q_1q_2 S_{3,i}- {5\choose 2} q_1^2 q_2^2  S_{1,i}= %\alpha_i ^5-5q_1q_2(\alpha_i^3- 3q_1q_2\alpha_i) -10q_1^2 q_2^2\alpha_i=
    \alpha ^5-5q_1q_2\alpha^3+ 5q_1^2q_2^2\alpha, %\\ &% \end{align*}\begin{align*}
S_{\alpha, 6} =%\defeq S_{3,i}^2- 2q_1^3q_2^3=(\alpha_i^3- 3q_1q_2\alpha_i)^2- 2q_1^3q_2^3=
   \alpha^6- 6q_1q_2\alpha^4  +9 q_1^2q_2^2\alpha^2- 2q_1^3q_2^3 \\   
   &S_{\alpha, 7} =%\defeq % \alpha_i^7- 7 q_1q_2 (\alpha_i ^5-5q_1q_2\alpha_i^3+ 5q_1^2q_2^2\alpha_i) - 21 q_1^2 q_2^2 (\alpha_i^3- 3q_1q_2\alpha_i) -35  q_1^3 q_2^3  \alpha_i =\\
    \alpha^7- 7 q_1q_2 \alpha ^5 +14 q_1^2q_2^2\alpha^3 -7 q_1^3q_2^3\alpha, \quad  \text{and, in general}\\  
  &S_{\alpha, k}=  \alpha_i^k- k q_1q_2 \alpha_i ^{k-2} +\frac{k(k-3)}{2} q_1^2q_2^2\alpha_i^{k-4} -(k^2-8k+14)q_1^3q_2^3\alpha_i^{k-6}+\cdots
   % & s_{k-4}=k-2+s_{k-5}=k-2+k-3+s_{k-6}=k-2+k-3+k-4+s_{k-7}=k^2-(2+3+4+\ldots +k+1)=k^2-(k+1)(k+2)/2-1    
    \end{align*}
  The coefficients of $S_{\alpha, k}, k\geq 8$ are to be similarly deduced recursively; 
\item Therefore, if $$ E_{k}\defeq \left\{\begin{matrix} \sum\limits _{\alpha\neq -(q_1+q_2)} \alpha^k=\sum\limits_{\lambda\neq 0}  \left(\lambda^2-(q_1+q_2)\right)^k, &k\geq 0\\ 0, &  k<0\end{matrix} \right., $$  then
%     $$%\sum\limits _{\alpha\neq q_1+q_2} (\xi_{\alpha,1}^3 +\xi_{\alpha,2}^3)=
%     \sum\limits _{\alpha\neq -(q_1+q_2)} S_{\alpha,3}=E_3- 3q_1q_2 E_1 , \quad %\sum\limits _{\alpha\neq q_1+q_2} (\xi_{\alpha,1}^4 +\xi_{\alpha,2}^4)=
%\sum\limits _{\alpha\neq q_1+q_2} S_{\alpha,4}=E_4- 4q_1q_2E_2+  2(\mu-1) q_1^2q_2^2 $$
%$$  \sum\limits _{\alpha\neq -(q_1+q_2)} S_{\alpha,5}=E_5- 5q_1q_2 E_3 + 5q_1^2q_2^2E_1, \quad %\sum\limits _{\alpha\neq q_1+q_2} (\xi_{\alpha,1}^4 +\xi_{\alpha,2}^4)=
%\sum\limits _{\alpha\neq q_1+q_2} S_{\alpha,6}=E_6- 6q_1q_2E_4+  9q_1^2q_2^2E_3-2(\mu-1)q_1^3q_2^3  $$%i.e., \begin{align*}
%$$  \sum\limits _{\alpha\neq -(q_1+q_2)} S_{\alpha,7}=E_7- 7q_1q_2 E_5 + 14q_1^2q_2^2E_3 -7 q_1^3q_2^3 E_1, $$
%\sum\limits _{\alpha\neq q_1+q_2} (\xi_{\alpha,1}^4 +\xi_{\alpha,2}^4)=
$$\sum\limits _{\alpha\neq -(q_1+q_2)} S_{\alpha,k}=E_k- kq_1q_2E_{k-2} +\frac{k(k-3)}{2} q_1^2q_2^2E_{k-4} -(k^2-8k+14)q_1^3q_2^3E_{k-6}+\cdots  $$%&S_{1,i}\defeq \lambda^2_i - (q_1+q_2) \\
%   &S_{2,i}\defeq S_{1,i}^2- 2q_1q_2=\lambda_i^4-2(q_1+q_2) \lambda_i^2  +(q_1^2+q_2^2)\\ 
%   &S_{3,i}\defeq S_{1,i}^3- 3q_1q_2 S_{1,i}=\lambda_i^6 - 3(q_1+q_2) \lambda_i^4 +3(q_1^2+q_2^2+ q_1q_2)\lambda_i^2 -(q_1+q_2) (q_1^2+q_2^2- q_1q_2)\\  
%   &S_{4,i}\defeq S_{2,i}^2- 2q_1^2q_2^2=\lambda_i^8 - 4(q_1+q_2)\lambda_i^6 +2 (3q_1^2+3q_2^2+ 4q_1q_2) \lambda_i^4- 4(q_1+q_2) (q_1^2+q_2^2)\lambda_i ^2 +(q_1^4+q_2^4)\\ 
%   &S_{5,i}\defeq S_{1,i}^5- {5\choose 1} q_1q_2 S_{3,i}- {5\choose 2} q_1^2 q_2^2  S_{1,i}\\
%   &S_{6,i}\defeq S_{3,i}^2- 2q_1^3q_2^3 \\   
%   &S_{7,i}\defeq S_{1,i}^7- {7\choose 1} q_1q_2 S_{5,i}- {7\choose 2} q_1^2 q_2^2  S_{3,i}- {7\choose 3} q_1^3 q_2^3  S_{1,i}, \\&  etc..     
%   \end{align*}
\end{enumerate}

\subsection{Main Result and Examples}Putting all these observations together, we obtain the following  formula for the number of $2k$-cycles that can be computed  recursively. %$$

\begin{theorem}\label{recursive} Let $G = (V = U\cup  W, E)$ be a connected bi-regular bipartite graph such that each node in $U$ has degree $q_1 +1$ and each node in $W $ has degree $q_2 +1$, where $q_2\geq 2$, $q_1 \geq 1$ and $q_2 \geq  q_1$. Also, assume that $|U| = n$ and $|W| = m$. 

Let $\{\alpha\} $ be the  set of all  %$\lambda_0=(q_1+1)(q_2+1)> \lambda_1>\ldots> \lambda_{l}, $ be the nonzero distinct eigenvalues of $HH^T$,  of multiplicities $\mu_0=1, \mu_1,  \cdots, \mu_{l}$, 
%$\alpha_0 , %= (q_1+1)(q_2+1)- (q_1+q_2),
%\alpha_1 %= \lambda_1- (q_1+q_2)
%,  \ldots,  \alpha_{l} %=\lambda_{l} - (q_1+q_2)
nonzero eigenvalues of $HH^T-(q_1+q_2)I_{m}$ and %of multiplicities $\mu_0=1, \mu_1,  \cdots, \mu_{l}$,
$\mu\defeq |\{\alpha \neq -(q_1+q_2)\}|  $  denote the rank of $H$ (which is equal to the rank of $HH^T$).  
Then, the number of cycles $N_{2k}$ in $G$ of length $2k$, $k\leq g$ is equal to \begin{align*} 
&N_{2k}=
\frac{|E| -(m+n) 
+ (-q_1)^k(n-\mu) +(-q_2)^k(m-\mu) + \sum\limits _{\alpha\neq -(q_1+q_2)} S_{\alpha,k}}{2k}\\
%
%N_{2k}= %&\frac{|E| -(m+n) +(-q_1)^k(n-\mu) +(-q_2)^k(m-\mu) + \sum\limits_{i=1}^l \mu_iS_{k,i} }{2k}=\\= 
%  &\frac{|E| -(m+n)+1 + (q_1q_2)^k+ (-q_1)^k(n-\mu) +(-q_2)^k(m-\mu) + \sum\limits_{i=1}^{l} \mu_iS_{k,i} }{2k} \\
 \end{align*}
where
\begin{align*}
  &  S_{\alpha, 1} =\alpha, \quad  S_{\alpha, 2} =S_{\alpha, 1}^2-2q_1q_2=\alpha^2-2q_1q_2, \quad 
    S_{\alpha, k}=%S_{k-1,i} S_{1,i}-q_1q_2 S_{k-2,i} =
     \alpha S_{\alpha, k-1}-q_1q_2 S_{\alpha, k-2}, k\geq 3.\end{align*} %and, therefore,
%  \begin{align*}   &
%    \sum\limits _{\alpha\neq -(q_1+q_2)}  S_{\alpha, 1} =\sum\limits _{\alpha\neq -(q_1+q_2)} \alpha= |E|-\mu(q_1+q_2), \quad \\& \sum\limits _{\alpha\neq -(q_1+q_2)}S_{\alpha, 2} =\sum\limits _{\alpha\neq -(q_1+q_2)}S_{\alpha, 1}^2-2\mu q_1q_2=\sum\limits _{\alpha\neq -(q_1+q_2)}\alpha^2-2q_1q_2\mu, \quad 
%    S_{\alpha, k}=%S_{k-1,i} S_{1,i}-q_1q_2 S_{k-2,i} =
%     \alpha S_{\alpha, k-1}-q_1q_2 S_{\alpha, k-2}, k\geq 3,     %  \sum\limits _{\alpha\neq -(q_1+q_2)} S_{\alpha,k}=E_k- kq_1q_2E_{k-2} +\frac{k(k-3)}{2} q_1^2q_2^2E_{k-4} -(k^2-8k+14)q_1^3q_2^3E_{k-6}+\cdots  
%\end{align*}
   %   S_{1,i}\defeq \alpha_i, & \quad 
\end{theorem} 

\begin{IEEEproof} The proof is given in Observations 1-8 listed in Section~\ref{sec:obs}. \end{IEEEproof} 
\begin{remark}
  For a QC-LDPC code with a $n_c\times n_v$ biregular parity-check matrix $P(x)$, modulo $x^N-1$, 
  $$n=n_vN, m=n_cN, |E|=n_cN(q_2+1)=n_vN(q_1+1),  \mu=n_cN-2.$$ If the matrix consists only of monomial and no zeros, then $q_1+1=n_c, q_2+1=n_v $. 
  \end{remark} 
  
   \begin{example} \label{example2} Let $H(x)$ be the following  $3\times n$ monomial matrix
    \begin{align*}&H(x)=\begin{bmatrix} 1&1&1&\ldots &1\\1&x&x^2&\ldots &x^{n-1}\\1&x^{i_1}&x^{i_2}&\ldots &x^{i_{n-1}} \end{bmatrix}, \end{align*}
modulo $x^n-1$, $n$ odd, for which the graph is known to have girth 6.\footnote{For $n$ odd, $N=n$ is the smallest lifting factor needed for a $3\times n$  protograph to achieve girth 6.}
From~\cite{sm22}, girth 6  is achieved with  
 \begin{align}\label{girth-condition} \{i_1,  \ldots i_{n-1}\}=\{1-i_1,  \ldots , n-1-i_{n-1}\} =\{1,  \ldots ,n-1\}.\end{align} 
 
We have $q_1=2, q_2=n-1.$ If  $f\defeq 1+x+\cdots+x^{n-1}$, then $f(\rho)=0$ for $\rho= e^{i2\pi j/n}, j\neq 0$, and so $f(\rho)=n$,  for $\rho=1$,  and  
% {\small \begin{align*} H(x)H(x)^T&=\begin{bmatrix} n&f&f\\f&n&f\\f&f&n\end{bmatrix} \Rightarrow 
%H(\rho)H^T(\rho)= \begin{bmatrix} n&0&0\\0&n&0\\0&0&n\end{bmatrix}, \rho= e^{i2\pi j/n}, j\neq 0,  
%H(\rho)H^T(\rho)= \begin{bmatrix} n&n&n\\n&n&n\\n&n&n\end{bmatrix},  \rho= 1. \end{align*}}
\begin{align*} H(x)H(x)^T&=\begin{bmatrix} n&f&f\\f&n&f\\f&f&n\end{bmatrix} \Rightarrow 
H(\rho)H^T(\rho)= \begin{cases} \begin{bmatrix} n&0&0\\0&n&0\\0&0&n\end{bmatrix}, &\rho= e^{i2\pi j/n}, j\neq 0,\\  
\begin{bmatrix} n&n&n\\n&n&n\\n&n&n\end{bmatrix},  &\rho= 1\end{cases} \end{align*}
Therefore, the corresponding scalar $3n\times 3n$  matrix $HH^\tr$ has nonzero eigenvalues $n$ of multiplicity $3(n-1)$ and $3n$ of multiplicity 1, with rank $3n-2$. 
Consequently,  the nonzero eigenvalues of $HH^T-(1+n)I_{3n}$ and their multiplicities  are $$\alpha _0=2n-1, \mu_0=1, \alpha_1=-1, \mu_1= 3(n-1).$$ 
Therefore, for $k=1,2,3,4,5$,   \begin{align*} 
N_{2k}=& %&\frac{|E| -(m+n) +(-q_1)^k(n-\mu) +(-q_2)^k(m-\mu) + \sum\limits_{i=1}^l m_iS_{k,i} }{k}=\\= 
%\frac{2n^2-3n +(n^2-3n+2)(-2)^k +2 (1-n)^k +%1 + (q_1q_2)^k
  %\sum\limits _{\alpha\neq -(q_1+q_2)} S_{\alpha,k}
%E_k- 2k(n-1)E_{k-2} +2k(k-3)(n-1)^2E_{k-4} %-(k^2-8k+14)8(n-1)^3E_{k-6}+\cdots
%}{2k} =\\&
\frac{2n^2-3n+ 1+(n^2-3n+2)(-2)^k +2 (1-n)^k +2^k(n-1)^k + 3(n-1)S_{\alpha_1,k}}{2k}=\\&(n-1)\cdot \frac{2n-1+(n-2)(-2)^k -2 (1-n)^{k-1} +2^k(n-1)^{k-1} + 3S_{\alpha_1, k}}{2k} \\%\end{align*}}
%\begin{align*} %&\frac{2n^2-3n +(n^2-3n+2)(-2)^k +2 (1-n)^k + 3(n-1)S_{k,1}+ S_{k,2}}{k}=\\
S_{\alpha_1, 1}=&  -1; %2n-1;\\
 S_{\alpha_1, 2} = 5-4n;  %4n^2-8n+5=1+4(n-1)^2;  \\ 
S_{\alpha_1, 3} = %\defeq S_{2,1} S_{1,1}-2(n-1) S_{1,1} =-(5-4n)+2(n-1)=
   6n-7; %8n^3-24n^2+24n-7=1+8(n-1)^3; \\  
 S_{\alpha_1, 4}= %\defeq S_{3,1} S_{1,1}-2(n-1) S_{2,1}= -(6n-7)-2(n-1)(5-4n)=  
   8n^2-24n+17; %16n^4-64n^3+96n^2-64n+17=1+16(n-1)^4; \\ 
S_{\alpha_1, 5} = %\defeq S_{4,1} S_{1,1}-2(n-1) S_{3,1}= -( 8n^2-24n+17)-2(n-1)(6n-7)  =
 -20n^2+50n-31. %\\& %(2n-1)^5- 10(n-1)(8n^3-24n^2+24n-7) - 40(n-1)^2(2n-1)
%32n^5-160n^4+320n^3-320n^2+160n-31=1+32(n-1)^5\\
  %&S_{6,i}\defeq S_{3,i}^2- 2q_1^3q_2^3 =(-15^3+24)^2 -2\cdot 8^3=11228177; 2\cdot 8^3=1024;   
  % &S_{7,i}\defeq S_{1,i}^7- {7\choose 1} q_1q_2 S_{5,i}- {7\choose 2} q_1^2 q_2^2  S_{3,i}- {7\choose 3} q_1^3 q_2^3  S_{1,i} ,  
   \end{align*}
    We obtain \begin{align*}   
   &N_2=%{2n^2-3n -2(n^2-3n+2) +2(1-n) - 3(n-1)+2n-1}{2} =
N_{4}=0, \quad %\frac{2n^2-3n +4(n^2-3n+2) +2 (1-n)^2 + 3(n-1)(5-4n)+ 4n^2-8n+5}{4}=0 ,  \\
N_6=n^2(n-1),  \quad 
N_8=\frac{3n^2(n-1)(3n-5)}{4},  \quad %= \frac{3}{4}(3n-5)N_6\\
N_{10}=3n^2(n-1)(n-2)^2. %= 3(n-2)^2N_6
  \end{align*}

For example, for $n=5$,  $H(x)=H_{n=5}(x)$ is a $3\times 5$ monomial matrix, modulo $x^5-1$, and for $n=7$,  $H(x)=H_{n=7}(x)$ is a $3\times 7$ monomial matrix, modulo $x^7-1$, 
$$H_{n=5}(x)=\begin{bmatrix} 1&1&1&1&1\\1&x&x^2&x^3&x^4\\1&x^4&x^3&x^2&x \end{bmatrix}, H_{n=7}(x)=\begin{bmatrix} 1&1&1&1&1&1&1\\1&x&x^2&x^3&x^4&x^5&x^6\\1&x^6&x^5&x^4&x^3&x^2&x \end{bmatrix}  . $$
We obtain, for $n=5$, 
\begin{align*} &N_7=N_{4}=0, \quad %{35 +12(-2) +2 (-4) + 12S_{1,1}+ S_{1,2}}{2} =0\\
%\frac{35 +12(-2)^2 +2 (-4)^2 + 12(-15)+ 65}{4}=0 ,  \\
N_6=25\cdot 4=100, \quad %\frac{35 +12(-2)^3 +2 (-4)^3 + 12S_{3,1}+ S_{3,2}}{3}=\frac{-189 +12\cdot(23)+513}{6}=100  \\ 
N_8= 750, \quad %\frac{35 +12(-2)^4 +2 (-4)^4 + 12S_{4,1}+ S_{4,2}}{8}=\frac{739+ 12\cdot97+ 4097}{8}=750\\
N_{10}=2700, %\frac{35 +12(-2)^5 +2 (-4)^5 + 12S_{5,1}+ S_{5,2}}{10}=\frac{-2397+12\cdot(-281)+ 32769}{10}=2700
\end{align*} 
and, for ${n=7}$, 
 \begin{align*} &N_2=N_{4}=0, %{35 +12(-2) +2 (-4) + 12S_{1,1}+ S_{1,2}}{2} =0\\
 %\frac{35 +12(-2)^2 +2 (-4)^2 + 12(-15)+ 65}{4}=0 ,  \\
N_6=49\cdot 6=294, %\frac{35 +12(-2)^3 +2 (-4)^3 + 12S_{3,1}+ S_{3,2}}{3}=\frac{-189 +12\cdot(23)+513}{6}=100  \\ 
N_8= \frac{3\cdot 49\cdot 6\cdot 16}{4}=3528, %\frac{35 +12(-2)^4 +2 (-4)^4 + 12S_{4,1}+ S_{4,2}}{8}=\frac{739+ 12\cdot97+ 4097}{8}=750\\
N_{10}=3\cdot 49\cdot 6\cdot 25=22050.  %\frac{35 +12(-2)^5 +2 (-4)^5 + 12S_{5,1}+ S_{5,2}}{10}=\frac{-2397+12\cdot(-281)+ 32769}{10}=2700
\end{align*} 

In general,  for a given $n$, no matter how we choose the exponents $i_1, i_2, \ldots, i_n$, and provided that~\eqref{girth-condition} is satisfied, 
the number of cycles $N_{2k}$ is the same for all such codes. This follows since all such matrices have the same eigenvalues and same values $q_1, q_2$. 
\end{example} 

\begin{remark} If we apply the original Theorem~\ref{thm2}, we would need to compute $\xi_1^k +\xi_2^k$, for each $k$, where $\xi_1, \xi_2$ are roots of a polynomial with integer coefficients, but themselves not necessarily integer. Also, a formula in terms of $n$ might not look as nice, since we need to use the binomial formula to obtain $\xi_1, \xi_2$, and hence, we would have square roots of expressions in $n$. The recursive formula from Theorem~\ref{recursive} allows us to use the integer coefficients directly and thus to obtain clean formulas in terms of positive powers of $n$. In addition, there are situations when  the eigenvalues are not integers but the characteristic polynomial has   integer coefficients. Again, the recursive formula and Newton's identities  can allow us to  directly use these coefficients,  and bypass even  the computation  of the eigenvalues of $HH^\tr$ and bypass  the approximations that are necessary when plugging them in in the original formulas (which might lead to being off slightly in the computation of $N_{2k}$). We exemplify this in the following example. 
 \end{remark}

\noindent{\bf Example \ref{Tannerexample} (cont.).} We revisit the $3\times 5$, $[155,64,20]$ code, $N=31$. The positive eigenvalues of $HH^\tr$ were found to be $\lambda_1 = 8.6801,\lambda_2 = 4.8459,  \lambda_3 = 1.4740$, each with multiplicity 30,  such that $\lambda_i, i=1,2,3,$ satisfy the equation $t^3 -15t^2 +62t-62=0$,  and $\lambda^2_0=15$ with multiplicity 1.\footnote{For notational simplicity, we will use $\lambda$ instead of $\lambda^2$ for the eigenvalues of $HH^\tr$.}  Therefore, $$\sum\limits_{i=1}^3 \lambda_i=15,\sum\limits_{{i, j=1}\atop {i< j}}^3\lambda_i\lambda_j=62, \prod\limits_{i=1}^3\lambda_i=62,$$ 
from which we obtain
%we have the following Newton's identities: 
\begin{align*} &E_1\defeq \sum_{i=1}^3 \alpha_i= \sum_{i=1}^3 \lambda_i -3(q_1+q_2)=-3, \\ 
&\sum_{{i, j=1}\atop {i< j}}^3 \alpha_i\alpha_j= \sum_{{i, j=1}\atop {i< j}}^3 \lambda_i \lambda_j - 2(q_1+q_2)\sum_{i=1}^3 \lambda_i +3(q_1+q_2)^2=-10,  \\
&\prod\limits_{i=1}^3\alpha_i =\prod\limits_{i=1}^3 (\lambda^2_i -(q_1+q_2) )=\prod\limits_{i=1}^3\lambda_i -(q_1+q_2)\sum_{{i, j=1}\atop {i< j}}^3 \lambda_i \lambda_j +(q_1+q_2)^2\sum_{i=1}^3 \lambda_i-(q_1+q_2)^3=14. %\\& 62-6\cdot 62+6^2\cdot 15-6^3=
\end{align*}
Note that we compute them as above instead of using the approximative values of the eigenvalues as to obtain integers. 

Therefore, the  characteristic polynomial of $H(\rho) H^\tr (\rho) -(q_1+q_2)I_{3}$ is  $\alpha^3 +3\alpha^2-10\alpha -14,$ where $\rho $ is the $31$st roots of unity. 
 Using Newton's identity for these three roots, we obtain 
{\small \begin{align*}& \sum\limits_{i=1}^3 \alpha_i=-3, \quad \sum\limits_{i=1}^3 \alpha^2_i=-3\sum\limits_{i=1}^3 \alpha_i- 2\cdot (-10)=29, 
\quad  \sum_{i=1}^3 \alpha^3_i=-3\sum\limits_{i=1}^3 \alpha^2_i+10\sum\limits_{i=1}^3 \alpha_i+3\cdot 14=% 29(-3)+10(-3)+42=
-75,\\
& \sum_{i=1}^3 \alpha^4_i=-3\sum\limits_{i=1}^3 \alpha^3_i+10\sum\limits_{i=1}^3 \alpha^2_i+14\sum\limits_{i=1}^3 \alpha_i=%(-3)(-75)+10(29)+14(-3)=
473, \quad 
 \sum_{i=1}^3 \alpha^5_i=-3\sum\limits_{i=1}^3 \alpha^4_i+10\sum\limits_{i=1}^3 \alpha^3_i+14\sum\limits_{i=1}^3 \alpha^2_i= %(-3)473+10(-75)+14(29)=
-1763,  \\ 
& \sum_{i=1}^3 \alpha^6_i=-3\sum\limits_{i=1}^3 \alpha^5_i+10\sum\limits_{i=1}^3 \alpha^4_i+14\sum\limits_{i=1}^3 \alpha^3_i= %(-3)(-1763)+10(473)+14(-75)=
8969, \quad  
\sum_{i=1}^3 \alpha^7_i=-3\sum\limits_{i=1}^3 \alpha^6_i+10\sum\limits_{i=1}^3 \alpha^5_i+14\sum\limits_{i=1}^3 \alpha^4_i= %(-3)8969+10(-1763)+14(473)=
-37915.\end{align*}}
We use each of these three $\alpha_i$ to recursively compute $S_{\alpha_i, k}$ and, in particular, the sum  $\sum\limits_{i=1}^3S_{\alpha_i, k}$. We then use it  with multiplicity 30 in the formula for $N_{2k}$, where
$$ N_{2k}= \frac{218+8^k +64(-2)^k+2(-4)^k +30\sum\limits_{i=1}^3S_{\alpha_i, k}}{2k},$$ 

$S_{\alpha, 1} =\alpha, \quad  S_{\alpha, 2} =S_{\alpha, 1}^2-2q_1q_2=\alpha^2-2q_1q_2, \quad 
    S_{\alpha, k}=%S_{k-1,i} S_{1,i}-q_1q_2 S_{k-2,i} =
     \alpha S_{\alpha, k-1}-q_1q_2 S_{\alpha, k-2}, k\geq 3,$
     \begin{align*}&  \sum\limits_{i=1}^3 S_{\alpha_i,1}=\sum_{i=1}^3 \alpha_i =-3,\quad
\sum\limits_{i=1}^3 S_{\alpha_i,2}=\sum\limits_{i=1}^3 \alpha_i^2- 6q_1q_2=%29-6(8)=
  -19, \\ 
  &  \sum\limits_{i=1}^3 S_{\alpha_i,3}= \sum\limits_{i=1}^3 \alpha_i^3- 3q_1q_2\sum\limits_{i=1}^3 \alpha_i=%-75-3\cdot 8(-3)=
  -3,\quad 
    \sum\limits_{i=1}^3 S_{\alpha_i,4}=  \sum\limits_{i=1}^3 \alpha_i^4- 4q_1q_2\sum\limits_{i=1}^3 \alpha_i^2 + 6q_1^2q_2^2  =%473-4(8)(29)+6(8)^2=
  -71\\
  &\sum\limits_{i=1}^3 S_{\alpha_i,5}=\sum\limits_{i=1}^3 \alpha_i^5-5q_1q_2\sum\limits_{i=1}^3 \alpha_i^3+ 5q_1^2q_2^2\sum\limits_{i=1}^3 \alpha_i =%-1763 -5(8)(-75)+5(8)^2(-3)=
  277, \\
&\sum\limits_{i=1}^3 S_{\alpha_i,6}=\sum\limits_{i=1}^3 \alpha_i^6- 6q_1q_2\sum\limits_{i=1}^3 \alpha_i^4  +9 q_1^2q_2^2\sum\limits_{i=1}^3 \alpha_i^2- 6q_1^3q_2^3 =%8969-48(473)+9*8^2(29)-6(8^3)=
-103 \\   
   &\sum\limits_{i=1}^3 S_{\alpha_i,7}=%\defeq % \alpha_i^7- 7 q_1q_2 (\alpha_i ^5-5q_1q_2\alpha_i^3+ 5q_1^2q_2^2\alpha_i) - 21 q_1^2 q_2^2 (\alpha_i^3- 3q_1q_2\alpha_i) -35  q_1^3 q_2^3  \alpha_i =\\
    \sum\limits_{i=1}^3 \alpha_i^7- 7 q_1q_2\sum\limits_{i=1}^3 \alpha_i^5 +14 q_1^2q_2^2\sum\limits_{i=1}^3 \alpha_i^3 -7 q_1^3q_2^3\sum\limits_{i=1}^3 \alpha_i=% -37915-7*8*(-1763) +14(8)^2(-75)-7*8^3(-3)=
    4365  \end{align*}   
   % We have the formula $$ N_{2k}= \frac{218+8^k +64(-2)^k+2(-4)^k +30\sum\limits_{i=1}^3S_{\alpha_i, k}}{2k} $$
Thus,  we obtain, as expected (matching other papers, using different methods and algorithms, e.g., \cite{db19})
\begin{align*} %&N_2 %=\frac{218+8  -128 -8+ 30(-3) }{2} =0\\
&N_4=0, %\frac{218+8^2 +64\cdot (-2)^2 +2(-4)^2 + 30\cdot (-19)}{4} =0\\\end{align*} \begin{align*}
N_6= 0,  %\frac{218+8^3+ 64(-2)^3 +2(-4)^3  + 30 (-3)}{6} =0 ,\\  %\frac{35 +12(-2)^3 +2 (-4)^3 + 12S_{3,1}+ S_{3,2}}{3}=\frac{-189 +12\cdot(23)+513}{6}=100  \\ 
N_8= %  \frac{218+8^4  +64(-2)^4 +2(-4)^4+ 30 (-71)}{8}=
465,  
N_{10}= 3720, %  \frac{218+8^5  +64(-2)^5 +2(-4)^5+ 30 (277)}{10}=3720 \\
N_{12}= 22630, % \frac{218+8^6  +64(-2)^6 +2(-4)^6+ 30 (-103)}{12}= 22630\\
N_{14}=  156240. %\frac{218+8^7  +64(-2)^7 +2(-4)^7+ 30 (4365)}{14}= 156240.
\end{align*}

% {\tiny$$ N_{8}=
% \frac{218+q_1^4q_2^4  +64q_1^4 +2q_2^4+ 30\sum\limits_{i=1}^3 \lambda_i^8 - 120(q_1+q_2)\sum\limits_{i=1}^3 \lambda_i^6 +60 (3q_1^2+3q_2^2+ 4q_1q_2) \sum\limits_{i=1}^3 \lambda_i^4- 120(q_1+q_2) (q_1^2+q_2^2)\sum\limits_{i=1}^3 \lambda_i^2  +90(q_1^4+q_2^4)}{8} $$} $$=  \frac{30330+ 30\sum\limits_{i=1}^3 \lambda_i^8 - 720\sum_{i=1}^3 \lambda_i^6 +5520 \sum\limits_{i=1}^3 \lambda_i^5- 14400 \sum\limits_{i=1}^3 \lambda_i^2  }{8}= $$ $$ \frac{30330+ 30(6233) - 720(771) +5520(101)- 14400 (15) }{8}=465.$$
  
\begin{remark} %In both our examples, the eigenvalues that are not equal to the largest eigenvalue are all equal, however, this is not true in general. What is true, is that, 
For $N$ odd, the eigenvalues of $H(\rho^j)H^\tr (\rho^j)$ are the same as the ones for $H(\rho^{N-j})H^\tr (\rho^{N-j})$, so there can be as many as $1+\frac{N-1}{2}$ distinct sets of $n_c$ eigenvalues; while for $N$ even, we also have $\frac{N-2}{2} $ pairs of equal  sets of $n_c$ eigenvalues, while  $\rho=\pm1$  give two extra distinct sets of $n_c$ eigenvalues.  For codes with a lot of structure,  like the code in  Example~\ref{example2} or the Tanner code of Example~\ref{Tannerexample}, the eigenvalues  have only a small number of distinct values due to the constraints satisfied by the exponents. 
\end{remark}
  \section{Variations on the Formula for $N_{2k}$}
  In this section, we explore different ways of expressing the number of cycles of length $2k$, $k<g$, make connections to the necessary and sufficient conditions to achieve a given girth, and provide examples.
\subsection{A Formula  for $N_{2k} $ in Terms of Powers of $\alpha$}\label{Formula-alpha} 
Below we solve the recurrence relation and obtain $N_{2k}$ as function of  $E_k$, with $$E_{k}\defeq \left\{\begin{matrix} \sum\limits _{\alpha\neq -(q_1+q_2)} \alpha^k=\sum\limits_{\lambda\neq 0}  \left(\lambda-(q_1+q_2)\right)^k, &k\geq 0\\ 0, &  k<0\end{matrix} \right.,$$ where $\alpha$ runs over all eigenvalues of $HH^\tr-(q_1+q_2)I_{n_cN}$ and $\lambda$ runs over all eigenvalues of $HH^\tr.$\footnote{For notational simplicity, we again use $\lambda$ instead of $\lambda^2$ for the eigenvalues of $HH^\tr$.}

 \begin{corollary}\label{Formula 2} The formula for $N_{2k}$, as defined in Theorem~\ref{recursive}, in terms of $E_k$ is {\tiny  \begin{align*} %\label{formula} 
 N_{2k}=&%&\frac{|E| -(m+n) +(-q_1)^k(n-\mu) +(-q_2)^k(m-\mu) + \sum\limits_{i=1}^l m_iS_{k,i} }{k}=\\= 
\frac{|E| -(m+n) %1 + (q_1q_2)^k
+ (-q_1)^k(n-\mu) +(-q_2)^k(m-\mu) + %\sum\limits _{\alpha\neq -(q_1+q_2)} S_{\alpha,k}
E_k- kq_1q_2E_{k-2} +\frac{k(k-3)}{2} q_1^2q_2^2E_{k-4} -(k^2-8k+14)q_1^3q_2^3E_{k-6}+\cdots}{2k}=\\
& \frac{|E|\left(1+\frac{(-q_1)^k-1}{q_1+1}+\frac{(-q_2)^k-1}{q_2+1}\right) %1 + (q_1q_2)^k
- \mu\left((-q_1)^k+(-q_2)^k\right)+ %\sum\limits _{\alpha\neq -(q_1+q_2)} S_{\alpha,k}
E_k- kq_1q_2E_{k-2} +\frac{k(k-3)}{2} q_1^2q_2^2E_{k-4} -(k^2-8k+14)q_1^3q_2^3E_{k-6}+\cdots}{2k} \end{align*}} The remaining coefficients are computed using the recurrences in Theorem~\ref{recursive}.  
\end{corollary} \begin{IEEEproof}We apply the recurrence to find the coefficients of $\alpha_i^k$ and then sum them over all ${\alpha\neq -(q_1+q_2)}$. \end{IEEEproof} 
%And, as in the Examples~\ref{Tannerexample} and \ref{example2},  if the eigenvalues have large multiplicity, we can compute $\sum \alpha^k$ over all distinct $\alpha$, and these can be written in terms of the coefficients of the characteristic polynomial.  
  \subsection{Formula for $N_{2k} $ in Terms of Powers of $\lambda$}  \label{Formula-lambda}  If we want to write $N_{2k}$ as function of the eigenvalues $\lambda$ of $HH^\tr$,
 we substitute $\alpha=\lambda-q_1-q_2$ in Theorem~\ref{recursive}.
Below, we work out the formulas for $N_{2k} $, for $k=2,3,4,5,6, 7$. In a similar way to these computations, we can also  express  $N_{2k}$, $k\geq 8$,   in terms of the powers of $\lambda$.  We leave these for the interested reader.

 The notation $$\sum\limits \lambda\defeq \sum\limits_{\lambda} \lambda=\sum\limits_{\lambda\neq 0}\lambda$$ means that $\lambda $ runs through all eigenvalues of $HH^\tr$ and, equivalently, through the nonzero eigenvalues.  Since the zero eigenvalues do not add anything to the sum, we might as well leave the sum over all eigenvalues.

\begin{corollary} \label{cor2} For $2\leq k\leq 7$, $N_{2k}$, as defined in Theorem~\ref{recursive}, is \begin{align*}& N_{2k} =\\&\frac{|E|\left(1+\frac{(-q_1)^k-1}{q_1+1}+\frac{(-q_2)^k-1}{q_2+1}-k\frac{(-q_1)^k-(-q_2)^k}{q_1-q_2}\right) +\sum \lambda^k -k(q_1+q_2)\sum \lambda^{k-1} + \sum\limits_{i=2}^k (-1)^i a_{k,k-i} \sum \lambda^{k-i}}{2k}, \\
%&N_{2k}=\\%&\frac{|E| -(m+n) +(-q_1)^k(n-\mu) +(-q_2)^k(m-\mu) + \sum\limits_{i=1}^l m_iS_{k,i} }{k}=\\= 
%&\frac{|E|\left(1+\frac{(-q_1)^k-1}{q_1+1}+\frac{(-q_2)^k-1}{q_2+1}-k\frac{(-q_1)^k-(-q_2)^k}{q_1-q_2}\right) +\sum \lambda^k -k(q_1+q_2)\sum \lambda^{k-1} %1 + (q_1q_2)^k
% + %\sum\limits _{\alpha\neq -(q_1+q_2)} S_{\alpha,k}
%E_7- 7q_1q_2E_{5} +14 q_1^2q_2^2E_{3} -7 q_1^3q_2^3E_{1}}{14}\\
%&\frac{|E|\left(1-\frac{q_1^7+1}{q_1+1}-\frac{q_2^7+1}{q_2+1} +7\frac{q_1^7-q_2^7}{q_1-q_2}\right) +\sum \lambda^7 
%+a_{14,6} \sum \lambda^6+a_{14,5} \sum \lambda^5
%+a_{14,4}\sum \lambda^4 + a_{14,3} \lambda^3+a_{14,2}\sum \lambda^2 }{14}
\end{align*}
%where 
%\[
%\sum \lambda^7
%+ a_{14,6} \sum \lambda^6
%+ a_{14,5} \sum \lambda^5
%+ a_{14,4} \sum \lambda^4
%+ a_{14,3} \sum \lambda^3
%+ a_{14,2} \sum \lambda^2
%\]
where $a_{k, k-i}=0 $, for all  $i>k$, otherwise, 
\begin{align*} 
 a_{k,k-i} &=
% &= -\binom{7}{1}(q_1 + q_2),\quad 
%a_{k,k-3} = \binom{7}{2}(q_1 + q_2)^2 - 7 q_1 q_2,\\
%a_{k,k-4} &= -\binom{7}{3}(q_1 + q_2)^3 - 7\binom{5}{1} q_1 q_2 (q_1 + q_2)\\
%a_{14,3} &= \binom{7}{4}(q_1 + q_2)^4 - 7  \binom{5}{2}q_1 q_2(q_1 + q_2)^2 + 14 q_1^2 q_2^2\\
%a_{14,2} &=
\binom{k}{i}(q_1 + q_2)^i -k\binom{k-2}{i-2}q_1 q_2(q_1 + q_2)^{i-2} +\frac{k(k-3)}{2}\binom{k-4}{i-4}q_1^2 q_2^2 (q_1 + q_2)^{i-4}
\\&\hspace{10mm}-(k^2-8k+14)\binom{k-6}{i-6}q_1^3q_2^3(q_1+q_2)^{i-6}, \end{align*}
for all $2\leq k\leq 7$, $2\leq i\leq k$,  and where any binomial coefficient with any negative entries is defined to be $0$. 
\end{corollary} 
\begin{IEEEproof} Expanding the formula in Corollary~\ref{Formula 2} and grouping the coefficients of $\lambda^{k-i}$, $  2\leq i\leq k$, we obtain
\begin{align*} 
a_{k,k-2} &= \binom{k}{2}(q_1 + q_2)^2 - k \binom{k-2}{0}q_1 q_2,\\
a_{k,k-3} &= \binom{k}{3}(q_1 + q_2)^3 - k\binom{k-2}{1} q_1 q_2 (q_1 + q_2),\\
a_{k,k-4} &= \binom{k}{4}(q_1 + q_2)^4 - k  \binom{k-2}{2}q_1 q_2(q_1 + q_2)^2 + \frac{k(k-3)}{2}\binom{k-4}{0} q_1^2 q_2^2,\\
a_{k,k-5} &= \binom{k}{5}(q_1 + q_2)^5 - k  \binom{k-2}{3}q_1 q_2(q_1 + q_2)^3 + \frac{k(k-3)}{2}\binom{k-4}{1} q_1^2 q_2^2(q_1+q_2),\\
a_{k,k-6} &= \binom{k}{6}(q_1 + q_2)^6 - k  \binom{k-2}{4}q_1 q_2(q_1 + q_2)^4 + \frac{k(k-3)}{2}\binom{k-4}{2} q_1^2 q_2^2(q_1+q_2)^2\\&\hspace{10mm}- (k^2-8k+14)\binom{k-6}{0}q_1^3q_2^3,\\
a_{k,k-7} &= \binom{k}{7}(q_1 + q_2)^7 - k  \binom{k-2}{5}q_1 q_2(q_1 + q_2)^5 + \frac{k(k-3)}{2}\binom{k-4}{3} q_1^2 q_2^2(q_1+q_2)^3\\&\hspace{10mm}- (k^2-8k+14)\binom{k-6}{1}q_1^3q_2^3(q_1+q_2),\end{align*}
which can be compactly written with the one formula above. 
\end{IEEEproof}

  \begin{corollary} \label{cor4} We obtain the following formulas  for the number  of cycles in any bi-regular bipartite graph $N_{2k}$ in terms of $\sum\lambda^k$ and $N_{2j}, 1<j<k$: 
  
   \begin{enumerate}
\item  For $ g\geq 4,$ \begin{align*} N_4=& \frac{\sum \lambda^2- |E|\cdot A_4}{4},\text{  where }A_4\defeq (q_1+q_2+1),\\
 N_6=&-2(q_1+q_2)N_4+  \frac{\sum\lambda^3-|E|\cdot A_6 }{6},\text{  where } A_6\defeq  \left(\left(q_1+q_2+1\right)^2+q_1q_2\right);\end{align*}

\item For $ g\geq 6,$ 
 \begin{align*} N_8=&-3(q_1+q_2)N_6+\frac{ \sum\lambda^4-|E|\cdot A_8 }{8},\text{  where }A_8\defeq \left((q_1+q_2+1)^3+q_1q_2(3q_1+3q_2 +2)\right),\\
N_{10}=&  -4(q_1+q_2)N_8  -3\left(2(q_1+q_2)^2+q_1q_2\right)N_6+\frac{\sum \lambda^5 -|E|\cdot A_{10}}{10}, \text{  where }\\
& \quad \quad  {A_{10}\defeq-1+\frac{q_1^5+1}{q_1+1}+\frac{q_2^5+1}{q_2+1} -5\frac{q_1^5-q_2^5}{q_1-q_2}+5(q_1+q_2)A_8-a_{5,3}A_6+a_{5,2}A_4}, \\ 
&\quad \quad {a_{5,3} =10(q_1+q_2)^2-5q_1q_2, \text{ and }a_{5,2} =10(q_1+q_2)^3-15q_1q_2(q_1+q_2);}\end{align*} 

\item For  $g\geq 8$, 
%{\small  \begin{align*}  &N_{12}= -5(q_1+q_2)N_{10}-2\left(5(q_1+q_2)^2+2q_1q_2\right)N_8 + \frac{\sum \lambda^6 -|E|A_{12}}{12}, \text{ where}\\
%  & \quad \quad A_{12}=-1-\frac{q_1^6-1}{q_1+1}-\frac{q_2^6-1}{q_2+1} +6\frac{q_1^6-q_2^6}{q_1-q_2} +6(q_1+q_2)A_{10}-a_{6,4}A_8+a_{6,3}A_6-a_{6,2}A_4.\\
%&\quad \quad  \quad \quad  a_{6,4}=   15(q_1 + q_2)^2 - 6 q_1 q_2, ~
%a_{6,3}=  
%20(q_1 + q_2)^3
%-  24q_1 q_2 (q_1 + q_2), ~\\
%&\quad \quad \quad \quad a_{6,2}=
%15(q_1 + q_2)^4
%- 36 q_1 q_2 (q_1 + q_2)^2
%+ 9 q_1^2 q_2^2 \\ 
%&N_{14}=-6(q_1+q_2)N_{12}-5\left(3(q_1+q_2)^2+q_1q_2\right)N_{10}-4(q_1+q_2)\left(5(q_1+q_2)^2+6q_1q_2\right)N_8 +\frac{\sum \lambda^7-|E|A_{14}}{14},\\
%&\quad \quad A_{14}=-1+\frac{q_1^7+1}{q_1+1}+\frac{q_2^7+1}{q_2+1} -7\frac{q_1^7-q_2^7}{q_1-q_2}+7(q_1+q_2)A_{12}-a_{7,5}A_{10}+a_{7,4}A_8-a_{7,3}A_6+a_{7,2}A_4\\
%&\quad \quad \quad \quad a_{7,5} = 21(q_1 + q_2)^2 - 7 q_1 q_2, 
%a_{7,4} = 35(q_1 + q_2)^3 - 35 q_1 q_2 (q_1 + q_2)\\
%&\quad \quad\quad \quad a_{7,3} = 35(q_1 + q_2)^4 - 70 q_1 q_2(q_1 + q_2)^2 + 14 q_1^2 q_2^2, \\
%& \quad \quad\quad \quad a_{7,2} = 21(q_1 + q_2)^5 - 70 q_1 q_2(q_1 + q_2)^3 + 42 q_1^2 q_2^2(q_1 + q_2)\end{align*}}
\begin{align*}  &N_{12}= -5(q_1+q_2)N_{10}-2\left(5(q_1+q_2)^2+2q_1q_2\right)N_8 + \frac{\sum \lambda^6 -|E|A_{12}}{12},\end{align*}
where
  \begin{align*}A_{12}&=-1-\frac{q_1^6-1}{q_1+1}-\frac{q_2^6-1}{q_2+1} +6\frac{q_1^6-q_2^6}{q_1-q_2} +6(q_1+q_2)A_{10}-a_{6,4}A_8+a_{6,3}A_6-a_{6,2}A_4,\\
a_{6,4} &=   15(q_1 + q_2)^2 - 6 q_1 q_2, ~
a_{6,3}=  
20(q_1 + q_2)^3
-  24q_1 q_2 (q_1 + q_2), ~\\
a_{6,2} &=
15(q_1 + q_2)^4
- 36 q_1 q_2 (q_1 + q_2)^2
+ 9 q_1^2 q_2^2, \\ 
N_{14}&=-6(q_1+q_2)N_{12}-5\left(3(q_1+q_2)^2+q_1q_2\right)N_{10}-4(q_1+q_2)\left(5(q_1+q_2)^2+6q_1q_2\right)N_8\\& \quad\quad+\frac{\sum \lambda^7-|E|A_{14}}{14},\\
A_{14}&=-1+\frac{q_1^7+1}{q_1+1}+\frac{q_2^7+1}{q_2+1} -7\frac{q_1^7-q_2^7}{q_1-q_2}+7(q_1+q_2)A_{12}-a_{7,5}A_{10}+a_{7,4}A_8-a_{7,3}A_6+a_{7,2}A_4,\\
a_{7,5} &= 21(q_1 + q_2)^2 - 7 q_1 q_2, 
\quad a_{7,4} = 35(q_1 + q_2)^3 - 35 q_1 q_2 (q_1 + q_2),\\
a_{7,3} &= 35(q_1 + q_2)^4 - 70 q_1 q_2(q_1 + q_2)^2 + 14 q_1^2 q_2^2, \\
a_{7,2} &= 21(q_1 + q_2)^5 - 70 q_1 q_2(q_1 + q_2)^3 + 42 q_1^2 q_2^2(q_1 + q_2).\end{align*}
\end{enumerate}
  \end{corollary} 
  \begin{corollary}\label{cor5} We obtain the necessary and sufficient conditions to achieve a certain girth in terms of eigenvalues of $HH^T$ as 
  $$ g\geq 2k \text{ iff } \sum \lambda^j= |E|A_{2j}, \text{ for all } 2\leq j\leq k,$$
   where $A_{2j}$ are defined in Corollary~\ref{cor4}. 
   \begin{IEEEproof} The proofs of Corollary~\ref{cor4} and Corollary~\ref{cor5} can be found in Appendix~\ref{appendixA}. \end{IEEEproof} %      \begin{align*}
%    &{g\geq 6 \text{ iff } \sum \lambda^2=Tr(HH^\tr)^2= |E|(q_1+q_2+1)}  \\
%  &{ g\geq 8 \text { iff } g\geq 6 \text{ and }\sum\lambda^3=Tr(HH^\tr)^3 = |E|\left(\left(q_1+q_2+1\right)^2+ q_1q_2\right)} \\
%   &{ g\geq 10 \text { iff }g\geq 8 \text{ and }\sum\lambda^4= Tr(HH^\tr)^4= |E|\left((q_1+q_2+1)^3+q_1q_2(3q_1+3q_2 +2)\right)}\\
%   &{ g\geq 12 \text { iff }g\geq 10 \text{ and }\sum \lambda^5 =Tr(HH^\tr)^5= A_{10}\cdot |E|}\\
%   &{ g\geq 14 \text { iff }g\geq 12 \text{ and }\sum \lambda^6 =Tr(HH^\tr)^6= A_{12}\cdot |E|} 
%     \end{align*}
  \end{corollary} 
\noindent{\bf Example \ref{example2} (cont.).} For the girth 6,  $(3,n)$-protograph-based Tanner graphs with lifting factor $N=n$, $n$ odd, we obtain the following, $$q_1=2, q_2=n-1, q_1+q_2=n+1, q_1+q_2+1=n+2, q_1q_2=2(n-1), |E|= 3n^2. $$ We recall that $HH^T$ has nonzero  eigenvalues $n$, with multiplicity $3(n-1)$, and $3n$,  with multiplicity $1$. Therefore, 

\begin{align*}N_4&=  \frac{3(n-1)n^2+ 9n^2- 3n^2(n+2)}{4}=0,\\
N_6&= \frac{3(n-1)n^3+ 27n^3-3n^2\left((n+2)^2+2(n-1)\right) }{6}=n^2(n-1),\\
 N_8&=-3(n+1)n^2(n-1)+\frac{ 3(n-1)n^4+ 81n^4-3n^2\left((n+2)^3+2(n-1)(3n+5)\right)}{8}
\\ &=\frac{3n^2(n-1)(3n-5)}{4},\\
N_{10}&=  -(n+1)3n^2(n-1)(3n-5)  -3\left(2(n+1)^2+2(n-1)\right)n^2(n-1)\\&\quad\quad+\frac{3(n-1)n^5+(3n)^5 -3n^2\cdot A_{10}}{10}. 
\end{align*}
We see that  $N_{10}$ is tedious to compute in this way, but was easy to compute with the recursive method we used earlier. This example demonstrates the comparative simplicity of our recursive formula, which is even computable by hand.
 %\end{example} 
 
 \begin{example}
 For any $(3,5)$-biregular graphs, we obtain $$q_1=2, q_2=4, q_1+q_2=6, q_1+q_2=7, q_1q_2=8.$$ 
 Then\begin{align*}&\text{for } g\geq 4 , N_4=
 \frac{\sum \lambda^2- 7|E|}{4}, \quad 
N_6=-12N_4+  \frac{\sum\lambda^3-57|E|}{6}, ~ A_4=7, ~A_6=57;\\
&\text{for } g\geq 6 , N_8=-18N_6+\frac{ \sum\lambda^4-503|E|
}{8}, A_8=503;\\
&\text{for } g\geq 6 ,  N_{10}=  -24N_8 -240N_6+\frac{\sum \lambda^5 
-4665|E|}{10}, ~A_{10}=4665;\\&{\text{for } g\geq 8,  N_{12}=-30N_{10}-392N_8 + \frac{ \sum \lambda^6 -44759|E|}{12}}, ~A_{12}= 44759;\\
&{\text{for } g\geq 8, N_{14}=-36N_{12}-580N_{10}-5472N_8 \frac{\sum \lambda^7-440217|E|}{14}, ~ A_{14}=440217}.\end{align*}
\end{example}

\noindent{\bf Example \ref{Tannerexample} (cont.).}  For the $(3,5)$ Tanner code,  the sum of the three distinct eigenvalues not equal to the largest
can be obtained using Newton's identities:
\begin{align*} &\sum_{i=1}^3   \lambda_i=15, 
 \sum_{i=1}^3   \lambda_i^2=
101, \sum_{i=1}^3   \lambda_i^3=
771, \sum_{i=1}^3   \lambda_i^4 =15\cdot 771-62\cdot 101+62\cdot 15=
 6233, \\&\sum_{i=1}^3   \lambda_i^5=15\cdot 6233-62\cdot 771+62\cdot 101=
 51955, \\&\sum_{i=1}^3   \lambda_i^6=15\cdot 51955 -62\cdot 6233+62\cdot 771=440681, \\
& \sum_{i=1}^3   \lambda_i^7=15\cdot 440681 -62\cdot 51955+62\cdot 6233=3775451  \end{align*}
and, in general, for $k\geq 4,$
$\sum\limits_{i=1}^3   \lambda_i^k=15\sum\limits_{i=1}^3   \lambda_i^{k-1}- 62\sum\limits_{i=1}^3   \lambda_i^{k-2}+62\sum\limits_{i=1}^3   \lambda_i^{k-3}.$
Then $\sum \lambda^k=15^k +(N-1)\sum\limits_{i=1}^3   \lambda_i^k =15^k +30\sum\limits_{i=1}^3   \lambda_i^k$, since $N=31$ in this example. Also $|E| =N(q_1+1)(q_2+1) =31\cdot 15$. Since $g=8$, we obtain
 \begin{align*}&N_4= \frac{101\cdot 30+15^2- 7\cdot 31\cdot 15}{4}=0, \quad 
N_6=  \frac{771\cdot30+15^3-57\cdot 31\cdot 15}{6} =0,\\
& N_8=\frac{ 6233\cdot 30+15^4-503\cdot 15\cdot 31}{8}=465,\\
& N_{10}= -24N_8 -240N_6+\frac{51955\cdot 30+15^5-4665\cdot 15\cdot 31}{10}=3720,\\
&  N_{12}=-30\cdot 3720-392\cdot 465 + \frac{ 440681\cdot 30+15^6 -44759\cdot 15\cdot 31}{12}= 22630, \\
&N_{14}=-36\cdot 22630 -580\cdot 3720-5472\cdot 465 +\frac{3775451\cdot 30 +15^7-440217\cdot 15\cdot 31}{14}=156240.
%
%N_{12}=-30N_{10}-328N_8 +\frac{\sum\lambda^6-52823(15)(31)}{12}\frac{-52823|E| +\sum \lambda^6 
%-36 \sum \lambda^5
%+492\sum \lambda^{4} -3168 \sum \lambda^{3} +9648\sum \lambda^{2}}{12}}\\
%&{\text{for } g\geq 8, N_{14}=
%\frac{53577|E| +\sum \lambda^7 
%-42\sum \lambda^6+700\sum \lambda^{5}- 5880\sum \lambda^{4} + 26096\sum \lambda^{3}- 148752  \sum \lambda^{2} }{14}}
\end{align*}
   \section{Conclusions}\label{sec:conc} In this paper, we explored new connections between the cycles in the graph of QC-LDPC codes and the eigenvalues of the corresponding adjacency matrix. Several observations about the eigenvalues and their multiplicities allowed us to write new formulas to enumerate the number of cycles of biregular bipartite graphs, specifically: a computationally simple recursive formula, a function of the powers of the eigenvalues of $HH^\tr-(q_1+q_2)I$, and one in terms of powers of the eigenvalues of $HH^\tr$. 
\appendices\label{sec:appendices}  
\section{Proof of Corollaries  \ref{cor4} and \ref{cor5}} \label{appendixA}
  \begin{IEEEproof} First, 
\begin{align*}N_{2}&=%&\frac{|E| -(m+n) +(-q_1)^k(n-\mu) +(-q_2)^k(m-\mu) + \sum\limits_{i=1}^l m_iS_{k,i} }{k}=\\= 
\frac{|E| -(m+n) %1 + (q_1q_2)^k
-q_1(n-\mu) -q_2(m-\mu) + %\sum\limits _{\alpha\neq -(q_1+q_2)} S_{\alpha,k}
E_1}{2}\\&=\frac{|E| -m(1+q_2)-n(1+q_1) +\mu(q_1+q_2)+ %\sum\limits _{\alpha\neq -(q_1+q_2)} S_{\alpha,k}
\sum\limits \lambda -\mu(q_1+q_2)}{2}=
%& \frac{|E| -(m+n) %1 + (q_1q_2)^k
%-q_1(n-\mu) -q_2(m-\mu) + %\sum\limits _{\alpha\neq -(q_1+q_2)} S_{\alpha,k}
%Tr(HH^T) -\mu(q_1+q_2)}{2}\\  =
%&\frac{|E| -(m+n) +(-q_1)^k(n-\mu) +(-q_2)^k(m-\mu) + \sum\limits_{i=1}^l m_iS_{k,i} }{k}=\\= 
%\frac{|E| -2|E| %1 + (q_1q_2)^k
 %\sum\limits _{\alpha\neq -(q_1+q_2)} S_{\alpha,k}
%+|E|}{2}
0,  \end{align*}   as expected,  
since $\sum \lambda= Tr(HH^\tr)= |E|=n(q_1+1)=m(q_2+1)$. Also%\Right arrow q_1=\frac{|E|}{n}-1, q_2=\frac{|E|}{m}-1 %&\text{ since } E_1= \sum\limits _{\alpha\neq -(q_1+q_2)} \alpha=\sum\limits _{\lambda\neq0} \lambda -\mu(q_1+q_2)=Tr(HH^T) -\mu(q_1+q_2)=m(q_2+1)-\mu(q_1+q_2),
 \begin{align*} N_{4}=&%&\frac{|E| -(m+n) +(-q_1)^k(n-\mu) +(-q_2)^k(m-\mu) + \sum\limits_{i=1}^l m_iS_{k,i} }{k}=\\= 
%\frac{|E| -(m+n) %1 + (q_1q_2)^k
%+ q_1^2(n-\mu) +q_2^2(m-\mu) + %\sum\limits _{\alpha\neq -(q_1+q_2)} S_{\alpha,k}
%E_2-2\mu q_1q_2}{4}%\frac{\sum \lambda^2- |E|(q_1+q_2+1)}{4}
\frac{|E|\left(1+\frac{q_1^2-1}{q_1+1} +\frac{q_2^2-1}{q_2+1}-2\frac{q_1^2-q_2^2}{q_1-q_2}\right)+\sum \lambda^2}{4}=\frac{\sum \lambda^2- |E|(q_1+q_2+1)}{4} =\frac{\sum \lambda^2- |E|A_4}{4},\end{align*}  where $A_4\defeq (q_1+q_2+1)$. We obtain that  \begin{align*} \sum \lambda^2=Tr(HH^\tr)^2= 4N_4+|E|(q_1+q_2+1) =4N_4+|E|A_4,\end{align*} 
and, therefore, $$g\geq 6 \text{  iff }
\sum \lambda^2=Tr(HH^\tr)^2= |E|A_4=|E|(q_1+q_2+1).$$
%Recall that $g\geq 6$ iff $HH^T\triangle I=0$, and now we obtain that this is equivalent to $Tr(HH^\tr)^2= |E|(q_1+q_2+1)$. 
%
%This can indeed be seen by computing  $Tr(HH^\tr)^2$ directly  $ =m(q_2+1)^2+ q_1(q_2+1)m= m(q_2+1)^2+ \sum_{i\neq j}  Tr( C_{i,j}C_{i,j}^\tr)$  and there is no 2 in  $C_{i,j}C_{i,j}^\tr$ for all $i\neq j$ iff we get $q_1(q-2+1)$, (Work this out properly, if of interest). 
%where $$HH^T= \begin{bmatrix}q_2+1 &C_{1,2}&\ldots &C_{1,n} \\ \vdots&\vdots& \ldots &\vdots\\ C_{n,1} &C_{n,2}&\ldots &q_2+1\end{bmatrix}  $$
% \end{remark}

Continuing, 
 \begin{align*} N_{6}=&%&\frac{|E| -(m+n) +(-q_1)^k(n-\mu) +(-q_2)^k(m-\mu) + \sum\limits_{i=1}^l m_iS_{k,i} }{k}=\\= 
%\frac{|E| -(m+n) %1 + (q_1q_2)^k
% -q_1^3(n-\mu) -q_2^3(m-\mu) + %\sum\limits _{\alpha\neq -(q_1+q_2)} S_{\alpha,k}
%E_3- 3q_1q_2E_{1} }{6}\\ =&
\frac{|E|\left(1-\frac{q_1^3+1}{q_1+1}-\frac{q_2^3+1}{q_2+1}+3\frac{q_1^3-q_2^3}{q_1-q_2}\right)+\sum\lambda^3 -3(q_1+q_2)\sum\lambda^2}{6}
\\=&-2(q_1+q_2)N_4+  \frac{\sum\lambda^3-|E|\left(\left(q_1+q_2+1\right)^2+q_1q_2\right) }{6}=
-2(q_1+q_2)N_4+  \frac{\sum\lambda^3-|E|A_6 }{6}
\end{align*} where $A_6\defeq \left(\left(q_1+q_2+1\right)^2+q_1q_2\right).$ We obtain 
\begin{align*}\sum\lambda^3 = 6N_6 +12(q_1+q_2)N_4+|E|\left(\left(q_1+q_2+1\right)^2+ q_1q_2\right) =6N_6 +12(q_1+q_2)N_4+|E|A_6.\end{align*}  
Therefore, $g\geq 8$ iff
\begin{equation*}\left\{\begin{aligned} & \sum \lambda^2=Tr(HH^\tr)^2= |E|A_4=|E|(q_1+q_2+1), \\&\sum\lambda^3=Tr(HH^\tr)^3 =|E|A_6= |E|\left(\left(q_1+q_2+1\right)^2+ q_1q_2\right).\end{aligned}\right.\end{equation*}

%\begin{example} Let $$H(x)=\begin{bmatrix} 1&1&1&1\\1&x&x^2&x^3\\1&x^3&x&x^2\end{bmatrix} $$ with $N=4$, which gives a girth 4 QC-LDPC code. The nonzero eigenvalues are $12, 2, 2, 4, 4, 4,4,4, 6, 6$. We apply the formulas to get $\sum \lambda^2= 19\cdot 16$ and $|E|(q_1+q_2+1)=18\cdot 16$, and therefore  $$N_4=\frac{\sum \lambda^2- |E|(q_1+q_2+1)}{4}=\frac{16}{4}=4$$ and $$N_6= -2(q_1+q_2)N_4+  \frac{\sum\lambda^3-|E|\left(\left(q_1+q_2+1\right)^2+q_1q_2\right) }{6}=40.$$ 
%\end{example} 
Next, \begin{align*}&N_{8}= 
%\frac{|E| -(m+n) %1 + (q_1q_2)^k
%+ q_1^4(n-\mu) +q_2^4(m-\mu) + %\sum\limits _{\alpha\neq -(q_1+q_2)} S_{\alpha,k}
%E_4- 4q_1q_2E_{2} +2 \mu q_1^2q_2^2 }{8} =\\%&\frac{ \sum\lambda^4-4(q_1+q_2) \sum\lambda^3+(6q_1^2+6q_2^2+8q_1q_2) \sum\lambda^2 +|E|\left(1-4(q_1^2+q_2^2)(q_1+q_2)+\frac{q_1^4-1}{q_1+1}+\frac{q_2^4-1}{q_2+1}\right)}{8}=\\
\frac{ |E|\left(1+\frac{q_1^4-1}{q_1+1}+\frac{q_2^4-1}{q_2+1}-4\frac{q_1^4-q_2^4}{q_1-q_2}\right)+ \sum\lambda^4-4(q_1+q_2) \sum\lambda^3+a_{4,2}\sum\lambda^2 }{8},\end{align*}  where $a_{4,2}=\left(\binom{4}{2}(q_1+q_2)^2-4q_1q_2\right).$ Since $\sum \lambda^2=|E|A_4=|E|(q_1+q_2+1) $, we can rewrite it as 
\begin{align*}&N_{8}= \frac{|E|\left(a_{4,2}(q_1+q_2+1) +1+\frac{q_1^4-1}{q_1+1}+\frac{q_2^4-1}{q_2+1}-4\frac{q_1^4-q_2^4}{q_1-q_2}\right)+ \sum\lambda^4-4(q_1+q_2) \sum\lambda^3}{8}\end{align*} 
and, since $\sum\lambda^3 = 6N_6 +|E|A_6$,
\begin{align*}N_{8}
%\frac{ \sum\lambda^4-24(q_1+q_2)N_6-|E|\left((q_1+q_2+1)^3+q_1q_2(3q_1+3q_2 +2)\right)}{8}\\
&=-3(q_1+q_2)N_6+\frac{ \sum\lambda^4-|E|A_8}{8}\Rightarrow \sum\lambda^4= 8N_8+24(q_1+q_2)N_6 %-8(9q_1^2+9q_2^2+16q_1q_2)N_4
 +|E|A_8,\end{align*} 
 where $$ A_8\defeq  (q_1+q_2+1)^3+q_1q_2(3q_1+3q_2+2).$$  Then, $g\geq 10$ iff
\begin{equation*}\left\{\begin{aligned}   & \sum \lambda^2=Tr(HH^\tr)^2=|E|A_4= |E|(q_1+q_2+1), \\ &\sum\lambda^3=Tr(HH^\tr)^3 =|E|A_6= |E|\left(\left(q_1+q_2+1\right)^2+ q_1q_2\right),\\&\sum\lambda^4= Tr(HH^\tr)^4=|E|A_8= |E|\left((q_1+q_2+1)^3+q_1q_2(3q_1+3q_2 +2)\right).\end{aligned}\right.
\end{equation*}

Next, \begin{align*}
& N_{10}=  
\frac{|E|\left(1-\frac{q_1^5+1}{q_1+1}-\frac{q_2^5+1}{q_2+1} +5\frac{q_1^5-q_2^5}{q_1-q_2}\right) +\sum \lambda^5 -5(q_1+q_2)\sum \lambda^4 + a_{5,3}\sum \lambda^3-a_{5,2}\sum \lambda^2}{10},  \end{align*}with\begin{align*}&a_{5,3} =\binom{5}{2}(q_1+q_2)^2-5q_1q_2, \quad 
a_{5,2} =\binom{5}{3}(q_1+q_2)^3-15q_1q_2(q_1+q_2).\end{align*}
Since $N_4=0$, per assumption, and  $$\sum \lambda^2= |E|A_4, \quad\sum\lambda^3 = 6N_6 +|E|A_6, \text{ and }\sum\lambda^4=8N_8+24(q_1+q_2)N_6 
 +|E|A_8
,$$ 
 we obtain that  \begin{align*}N_{10}= & -4(q_1+q_2)N_8 -3\left(2(q_1+q_2)^2+q_1q_2\right)N_6+\frac{\sum \lambda^5 %40N_4(11q_1^3+11q_2^3+31q_1^2q_2+31q_1q_2^2)
-|E|\cdot A_{10}}{10},\\
%$$A=-1+\frac{q_1^5+1}{q_1+1}+\frac{q_2^5+1}{q_2+1} +5q_1^3+5q_2^3+5q_1^2+5q_2^2+20q_1^2q_2+20q_1q_2^2+10q_1^3q_2+10q_1q_2^3+20q_1^2q_2^2.$$
 \sum \lambda^5 = &10N_{10} +40(q_1+q_2)N_8 +30\left(2(q_1+q_2)^2+q_1q_2\right)N_6+|E|\cdot A_{10},%40N_4(11q_1^3+11q_2^3+31q_1^2q_2+31q_1q_2^2)
\end{align*} 
where $A_{10}\defeq-1+\frac{q_1^5+1}{q_1+1}+\frac{q_2^5+1}{q_2+1} -5\frac{q_1^5-q_2^5}{q_1-q_2}+5(q_1+q_2)A_8-a_{5,3}A_6+a_{5,2}A_4.$ Therefore, $g\geq 12$ iff\begin{equation*}\left\{\begin{aligned}  &
\sum \lambda^2=Tr(HH^\tr)^2= |E|A_4=|E|(q_1+q_2+1), \\ &\sum\lambda^3=Tr(HH^\tr)^3=|E|A_6 = |E|\left(\left(q_1+q_2+1\right)^2+ q_1q_2\right),\\
&\sum\lambda^4= Tr(HH^\tr)^4=|E|A_8= |E|\left((q_1+q_2+1)^3+q_1q_2(3q_1+3q_2 +2)\right),\\&\sum \lambda^5 =Tr(HH^\tr)^5=|E|A_{10}.\end{aligned}\right.\end{equation*}

Next, \begin{align*}N_{12}=%&\frac{|E| -(m+n) +(-q_1)^k(n-\mu) +(-q_2)^k(m-\mu) + \sum\limits_{i=1}^l m_iS_{k,i} }{k}=\\= 
%\frac{|E|\left(1+\frac{q_1^6-1}{q_1+1}+\frac{q_2^6-1}{q_2+1}\right) %1 + (q_1q_2)^k
%- \mu\left(q_1^6+q_2^6\right) %1 + (q_1q_2)^k
%+ %\sum\limits _{\alpha\neq -(q_1+q_2)} S_{\alpha,k}
%E_6- 6q_1q_2E_{4} +9q_1^2q_2^2E_{2} -2\mu q_1^3q_2^3 }{12}= \\
&\frac{|E|\left(1+\frac{q_1^6-1}{q_1+1}+\frac{q_2^6-1}{q_2+1} -6\frac{q_1^6-q_2^6}{q_1-q_2}\right) +\sum \lambda^6 
-6(q_1+q_2) \sum \lambda^5
+a_{6,4}\sum \lambda^4 - a_{6,3} \lambda^3+a_{6,2}\sum \lambda^2 }{12},\end{align*}
with
\begin{align*}& a_{6,4}=   \binom{6}{2} (q_1 + q_2)^2 - 6 q_1 q_2, \quad 
a_{6,3}=  
\binom{6}{3} (q_1 + q_2)^3
-  \binom{6}{1} \binom{4}{1}q_1 q_2 (q_1 + q_2), \\&
a_{6,2}=
\binom{6}{2} (q_1 + q_2)^4
- \binom{6}{1} \binom{4}{2} q_1 q_2 (q_1 + q_2)^2
+ 9 q_1^2 q_2^2.
\end{align*}
Equivalently,
\begin{align*}N_{12}&= -5(q_1+q_2)N_{10}-2\left(5(q_1+q_2)^2+2q_1q_2\right)N_8 + \frac{\sum \lambda^6 -|E|A_{12}}{12},\end{align*}
where $$A_{12}=-1-\frac{q_1^6-1}{q_1+1}-\frac{q_2^6-1}{q_2+1} +6\frac{q_1^6-q_2^6}{q_1-q_2} +6(q_1+q_2)A_{10}-a_{6,4}A_8+a_{6,3}A_6-a_{6,2}A_4.$$
We obtain that $$\sum \lambda^6 =12N_{12}+60(q_1+q_2)N_{10}+24\left(5(q_1+q_2)^2+2q_1q_2\right)N_8 +|E|A_{12}.$$

Finally, \begin{align*} N_{14}&=%&\frac{|E| -(m+n) +(-q_1)^k(n-\mu) +(-q_2)^k(m-\mu) + \sum\limits_{i=1}^l m_iS_{k,i} }{k}=\\= 
\frac{|E|\left(1-\frac{q_1^7+1}{q_1+1}-\frac{q_2^7+1}{q_2+1} +7\frac{q_1^7-q_2^7}{q_1-q_2}\right) +\sum \lambda^7 
-7(q_1+q_2) \sum \lambda^6+\sum\limits_{i=2}^{5}(-1)^ia_{7,7-i}\sum \lambda^{7-i} }{14}\\&=-6(q_1+q_2)N_{12}-5\left(3(q_1+q_2)^2+q_1q_2\right)N_{10}-4(q_1+q_2)[5(q_1+q_2)^2+6q_1q_2]N_8 \\&\quad\quad+\frac{\sum \lambda^7-|E|A_{14}}{14},\end{align*}
where $$A_{14}=-1+\frac{q_1^7+1}{q_1+1}+\frac{q_2^7+1}{q_2+1} -7\frac{q_1^7-q_2^7}{q_1-q_2}+7(q_1+q_2)A_{12}-a_{7,5}A_{10}+a_{7,4}A_8-a_{7,3}A_6+a_{7,2}A_4,$$
and \begin{align*} 
a_{7,5} &= \binom{7}{2}(q_1 + q_2)^2 - 7 q_1 q_2, \quad
a_{7,4} = \binom{7}{3}(q_1 + q_2)^3 - 7\binom{5}{1} q_1 q_2 (q_1 + q_2),\\
a_{7,3} &= \binom{7}{4}(q_1 + q_2)^4 - 7  \binom{5}{2}q_1 q_2(q_1 + q_2)^2 + 14 q_1^2 q_2^2,\\
a_{7,2} &= \binom{7}{5}(q_1 + q_2)^5 - 7  \binom{5}{3}q_1 q_2(q_1 + q_2)^3 + 14 q_1^2 q_2^2 \binom{3}{1}(q_1 + q_2).
\end{align*}
We obtain that \begin{align*}\sum \lambda^7 &=14N_{14}+84(q_1+q_2)N_{12}+70\left(3(q_1+q_2)^2-q_1q_2\right)N_{10}\\&\quad\quad-56(q_1+q_2)[5(q_1+q_2)^2-6q_1q_2]N_8 +|E|A_{14}.\end{align*}
%{\small 
%\begin{align*} %N_{12}=&%&\frac{|E| -(m+n) +(-q_1)^k(n-\mu) +(-q_2)^k(m-\mu) + \sum\limits_{i=1}^l m_iS_{k,i} }{k}=\\= 
%\frac{|E|\left(1+\frac{q_1^6-1}{q_1+1}+\frac{q_2^6-1}{q_2+1}\right) %1 + (q_1q_2)^k
%- \mu\left(q_1^6+q_2^6\right) %1 + (q_1q_2)^k
%+ %\sum\limits _{\alpha\neq -(q_1+q_2)} S_{\alpha,k}
%E_6- 6q_1q_2E_{4} +9q_1^2q_2^2E_{2} -2\mu q_1^3q_2^3 }{12}\\
%\end{align*} }
\end{IEEEproof} 
%  \frac{|E|\left(1+\frac{(-q_1)^k-1}{q_1+1}+\frac{(-q_2)^k-1}{q_2+1}\right) %1 + (q_1q_2)^k
%- \mu\left((-q_1)^k+(-q_2)^k\right)+ %\sum\limits _{\alpha\neq -(q_1+q_2)} S_{\alpha,k}
%E_k- kq_1q_2E_{k-2} +\frac{k(k-3)}{2} q_1^2q_2^2E_{k-4} -(k^2-8k+14)q_1^3q_2^3E_{k-6}+\cdots}{2k} 
%We note that a pattern is observed: 
%% \begin{itemize} \item for $k$ odd, 
%% $$N_{2k} =\frac{|E|\left(1-\frac{q_1^k+1}{q_1+1}-\frac{q_2^k+1}{q_2+1} +k\frac{q_1^k-q_2^k}{q_1-q_2}\right) +\sum \lambda^k -k(q_1+q_2)\sum \lambda^{k-1} +...}{2k} $$
%%\item $k$ even  $$N_{2k} =\frac{|E|\left(1+\frac{q_1^k-1}{q_1+1}+\frac{q_2^k+1}{q_2+1} -k\frac{q_1^k-q_2^k}{q_1-q_2}\right) +\sum \lambda^k -k(q_1+q_2)\sum \lambda^{k-1} +...}{2k} $$
%for all $k$, 
%$$N_{2k} =\frac{|E|\left(1+\frac{(-q_1)^k-1}{q_1+1}+\frac{(-q_2)^k-1}{q_2+1}-k\frac{(-q_1)^k-(-q_2)^k}{q_1-q_2}\right) +\sum \lambda^k -k(q_1+q_2)\sum \lambda^{k-1} +....}{2k} $$
%\end{itemize} 
%\end{appendix}

\bibliographystyle{IEEEtran}
\bibliography{roxanaeigen}
\end{document}